\documentclass[a4paper,preprint,aps]{revtex4}

\usepackage{graphicx}
\usepackage{dcolumn}
\usepackage{bm}

\newcommand{\eqa}{\begin{equation}}
\newcommand{\eqz}{\end{equation}}
\newcommand{\eqma}{\begin{eqnarray}}
\newcommand{\eqmz}{\end{eqnarray}}

\begin{document}
\newcommand{\e}{{\em e}~}
\title{W4 theory for computational thermochemistry: in pursuit of confident sub-kJ/mol predictions}
\author{Amir Karton, Elena Rabinovich, and Jan M. L. Martin*}
\affiliation{Department of Organic Chemistry, 
Weizmann Institute of Science, 
IL-76100 Re\d{h}ovot, Israel}
\email{comartin@wicc.weizmann.ac.il}

\author{Branko Ruscic}
\affiliation{Chemistry Division,
Argonne National Laboratory,
Argonne, IL 60439, USA}
\date{{\em Journal of Chemical Physics} MS\# {\bf A6.07.059}; Received July 7, 2006; Accepted August 10, 2006}

\begin{abstract}
In an attempt to improve on our earlier W3 theory [J. Chem. Phys. {\bf 120}, 4129 (2004)] we consider such refinements as more accurate estimates for the contribution of connected quadruple excitations ($\hat{T}_4$), inclusion of connected quintuple excitations ($\hat{T}_5$), diagonal Born-Oppenheimer
corrections (DBOC), and improved basis set extrapolation procedures. 
Revised experimental data for validation purposes were obtained from
the latest version of the ATcT (Active Thermochemical Tables) Thermochemical Network.
The recent CCSDT(Q) method offers a cost-effective way of estimating $\hat{T}_4$, but is insufficient by itself if the molecule exhibits some nondynamical
correlation. The latter considerably slows down basis set convergence for $\hat{T}_4$, and anomalous basis set convergence in highly polar systems makes 
two-point extrapolation procedures unusable. However, we found that the CCSDTQ$-$CCSDT(Q) difference converges quite rapidly with the basis set, and that the formula 1.10[CCSDT(Q)/cc-pVTZ+CCSDTQ/cc-pVDZ$-$CCSDT(Q)/cc-pVDZ] offers a very reliable as well as fairly cost-effective estimate of the basis set limit $\hat{T}_4$ contribution. The $\hat{T}_5$ contribution converges very rapidly with the basis set, and even a simple double-zeta basis set appears to be adequate. The largest $\hat{T}_5$ contribution found in the present work is on the order of 0.5 kcal/mol (for ozone). DBOC corrections are significant at the 0.1 kcal/mol level in hydride systems. Post-CCSD(T) contributions to the core-valence correlation energy are only significant at that level in systems with severe nondynamical correlation effects.
Based on the accumulated experience, a new computational thermochemistry
protocol for first-and second-row main-group systems, to be known as W4 theory, is proposed. Its computational cost is 
not insurmountably higher than that of the earlier W3 theory, while performance
is markedly superior. 
Our W4 atomization energies for a number of key species are 
in excellent agreement (better than 0.1 kcal/mol on average, 95\% confidence intervals narrower than 1 kJ/mol) with 
the latest experimental data obtained from Active Thermochemical Tables. Lower-cost variants are proposed: the sequence W1$\rightarrow$W2.2$\rightarrow$W3.2$\rightarrow$W4lite$\rightarrow$W4 is proposed as a converging hierarchy of computational thermochemistry methods.
A simple {\em a priori} estimate for the importance of post-CCSD(T) correlation contributions (and hence a pessimistic estimate for the error in a W2-type calculation) is proposed.
\end{abstract}
\maketitle
\section{Introduction}

In the past fifteen years, computational thermochemistry has matured to the point where its accuracy is often competitive with all but the most accurate experimental techniques.

A compact overview of computational thermochemistry methods in all their variety has very recently been published by one of us\cite{ARCC2006}, while a book with more detailed reviews of the various techniques was published in 2001\cite{CioslowskiBook}. In terms of `ready-made' nonempirical small-molecule methods of sub-kcal/mol accuracy, there have been two major developments in the last few years. One is the W$n$ family of computational thermochemistry protocols (to be discussed below)\cite{w1,w1eval,w3}, the
other has been the HEAT (Highly accurate Extrapolated Ab initio Thermochemistry) project by a multinational group of researchers\cite{HEAT}. In this context, mention should be made of the related `focal point approach' pioneered by Allen\cite{Allen} --- which is however more a general strategy than a precisely defined computational protocol --- as well as of the configuration interaction extrapolation based work of Bytautas and Ruedenberg\cite{Rue2004}.

The `W$n$ theory' naming scheme was introduced by analogy to the `G$n$ theory' family of methods of the late lamented Pople and coworkers\cite{g3x}.
The basic philosophy of the W$n$ family of methods can be outlined as follows:
\begin{itemize}
\setlength{\itemsep}{0pt}
\item All terms in the Hamiltonian that can reasonably contribute at the kJ/mol level to the atomization energy should be retained;
\item Basis set convergence is established for each contribution individually, and the smallest basis sets are used for each that still lead to acceptable basis set incompleteness errors for the relevant contribution;
\item As a result, computational effort is kept down to the minimum consistent with the required accuracy;
\item No parameters derived from experiment are employed: where possible, physically or empirically rational basis set extrapolations are employed.
\end{itemize}
W1 theory\cite{w1,w1eval} uses basis sets of no larger than $s$$p$$d$$f$$g$ quality, and no electron correlation methods more elaborate than CCSD(T). It represents an approximation to the relativistic, clamped-nuclei, basis set limit CCSD(T) energy. (Scalar relativistic corrections were obtained as one-electron Darwin and mass-velocity terms from averaged coupled pair wavefunctions, although the implementation in Gaussian 03\cite{g03} employs Douglas-Kroll scalar 
relativistics\cite{Dou74,Hes86} at the CCSD(T) level.) 
W2 theory\cite{w1,w1eval} aims at the same target as W1 theory but uses more elaborate $s$$p$$d$$f$$g$$h$ basis set and is in general more accurate. For systems dominated by a single reference determinant, W2 theory can usually achieve kJ/mol accuracy.

For systems with significant nondynamical correlation, the CCSD(T) limit differs significantly from the FCI (full configuration interaction) limit. Even for systems like the diatomics N$_2$, O$_2$, and F$_2$, W2 will be in error by 0.5--0.7 kcal/mol, and for ozone an error of 3 kcal/mol is seen\cite{w3}. 
Two main improvements were introduced in W3 theory. The first was a more robust scalar relativistic correction based on DKH-CCSD(T) calculations (Douglas-Kroll-Hess\cite{Dou74,Hes86}), with an eye to future extension of applicability to elements heavier than Ar. The second, which proved crucial, was an account for higher-order $\hat{T}_3$ effects --- i.e., the CCSDT$-$CCSD(T) difference --- on the one hand, and for connected quadruple excitations $\hat{T}_4$ on the other hand.
W3 proved much more robust to nondynamical correlation effects than its predecessors\cite{w3}: for systems dominated by a single reference determinant, it is comparable in accuracy to W2 theory.

Attempts to surpass W3-level accuracy (RMSD of about 1.2 kJ/mol, 95\% confidence interval of about 2.5 kJ/mol) were impeded by a number of problems, of which we shall cite only the two most important ones. 
On the one hand, it would definitely be necessary to consider connected quadruple excitation effects with basis sets larger than $spd$ quality, but this was precluded by the limiting $n^4N^6$ computational cost scaling (where $n$ and $N$ are the numbers of electrons and basis functions, respectively) of CCSDTQ calculations. 
On the other hand, we were working in an accuracy regime comparable to that of all but the very best experimental thermochemical data, and meaningful comparisons with experiment were just not possible beyond the kJ/mol range, except for a few select molecules.

The recent development of the CCSDT(Q) method\cite{Bomble2005} and of a code for arbitrary quasiperturbative techniques\cite{KallayGauss} opened a potential avenue for more cost-effective treatments of $\hat{T}_4$ on the theoretical front. In terms of availability of accurate and reliable thermochemical benchmarks, the advent of Active Thermochemical Tables afforded relief.

Active Thermochemical Tables (ATcT) are a new paradigm of how to obtain
accurate, reliable, and internally consistent  thermochemical values by
using all available knowledge\cite{Branko1,Branko2,Branko3,Branko4}, and overcome the
limitations that are deeply engrained in the traditional approach to
thermochemistry, such as that used in all traditional thermochemical
compilations.  As opposed to the traditional sequential approach, ATcT
derives its results from a Thermochemical Network (TN). 
Where available, the thermochemical values used in the present work for the purpose of benchmarking the W4 method have been obtained from the latest version of the Core (Argonne) Thermochemical Network, C(A)TN, that is currently under development and encompasses $\sim$700 chemical species containing H, O, C, N, and halogens, which are interlinked by $>$7000 thermochemically-relevant determinations\cite{Branko5}. In addition, the benchmark ATcT values for three sulfur-containing species have been obtained from a separate adjunct ATcT TN. The adjunct TN is currently under intense development and scrutiny as part of a joint Argonne-NIST project\cite{Branko5b}, which is attempting to address and resolve some of the inherent inconsistencies surrounding the accepted experimental thermochemical values for several key sulfur-containing species. Consequently, within the context of the present work, we have limited ourselves to using as benchmarks only three sulfur-containing species from ATcT (H$_2$S, SO and SO$_2$), whose values happen to be invariant (within 0.05 kcal/mol or better) to the finer details of the adjunct sulfur-containing ATcT TN.
For the  species of interest here, the
current version of C(A)TN includes all available experimental results
and also considers a  selection of prior highly accurate theoretical
results (with weights proportional to the expected uncertainties),
but, in order to keep the ATcT benchmarks used here independent of the current computational results, does not include W4. These will be added to the Thermochemical Network in a subsequent revision of C(A)TN. Full details of how the ATcT values were developed and what data are they based on will be published separately in a forthcoming series of papers\cite{Branko6}.

In the present work, we will:
\begin{itemize}
\setlength{\itemsep}{0pt}
\item explore basis set convergence of connected quadruple and quintuple
excitations in greater detail;
\item consider the effect of diagonal Born-Oppenheimer corrections;
\item present a computational protocol called W4 theory which yields significant improvements over W3 theory, for a tractable additional computational cost;
\item validate it against the best available benchmark values from ATcT;
\item present a simple energy-based diagnostic for the reliability of thermochemical approaches that neglect post-CCSD(T) correlation.
\end{itemize}

\section{Computational Details}

All quantum mechanical calculations were carried out on the Linux cluster of the Martin group at the Weizmann Institute. In practice, most results were obtained from four dual-processor 32-bit (Intel Xeon 2.8 and 3.06 GHz) and a single four-processor 64-bit (AMD Opteron 846) machine, all custom-built by Access Technologies of Re\d{h}ovot, Israel. These machines are equipped with 4-way and 8-way striped disk arrays, respectively, made up of 72 GB Ultra320 SCSI disks. The high sustained I/O throughputs of these machines\cite{iozone} proved essential to handle the daunting I/O requirements for many of the calculations reported.

All CCSD and CCSD(T) calculations were carried out using MOLPRO 2002.6\cite{molpro}. Conventional, rather than direct, algorithms were used throughout as this proved more efficient with the special I/O hardware available.

The DBOC calculations, and some full CCSDT calculations, were carried out using the Austin-Mainz-Budapest version of ACES II\cite{aces2de}. The remaining post-CCSD(T) calculations --- CCSDT(Q), CCSDTQ, CCSDTQ(5), CCSDTQ5, CCSDTQ5(6), CCSDTQ56, and full CI --- were carried out by means of the general coupled cluster code MRCC of K\'allay and coworkers\cite{mrcc}. The required integrals and SCF orbitals for the latter were obtained using ACES II.

The basis sets used are all of the correlation consistent family of Dunning and coworkers\cite{Dun89}. For the large-scale CCSD(T) calculation, we 
combined regular cc-pV$n$Z basis sets ($n$=D,T,Q,5,6) on hydrogen with aug-cc-pV$n$Z ([diffuse-function] augmented polarization consistent) basis sets\cite{Ken92} on boron through fluorine, and Wilson's aug-cc-pV$(n+d)$Z basis sets\cite{Wilson} on phosphorus through chlorine. The latter contain high-exponent $d$ functions to cope with `inner polarization' effects\cite{so2}: these occur in second-row elements in high oxidation states as a result of back-bonding into their low-lying $d$ orbitals\cite{cl2o7tae}. 
For post-CCSD(T) calculations, regular cc-pV$n$Z basis sets were employed unless
indicated otherwise.
In core-valence correlation calculations, we employed Peterson's core-valence weighted correlation consistent basis sets\cite{pwCVnZ}, aug-cc-pwCV$n$Z. Finally, the scalar relativistic calculations were carried out using both the unpublished Douglas-Kroll optimized correlation consistent basis sets by Oren and Martin\cite{oren-martin} and the 
PNNL (Pacific Northwest National Laboratory) relativistically contracted correlation consistent basis sets\cite{Dyall}. 
(We verified that fundamentally the same results were obtained.)

All reference geometries are obtained at the CCSD(T)/cc-pV(Q+d)Z level. 
Complete sets of geometries and total energies are available as supporting information to the present paper\cite{e-paps}.

\section{Results and discussion}

\subsection{Basis set convergence of post-CCSDT correlation contributions}

In order to get a clearer picture of basis set convergence behavior in post-CCSDT contributions, we considered in detail a set of 22 mostly diatomic molecules, namely, HF, H$_2$O, NH$_3$, PH$_3$, H$_2$S, HCl, CO, CS, Cl$_2$, ClF, N$_2$, B$_2$, CH, CN, NO, SO, O$_2$, F$_2$, C$_2$, BN, BeO, and MgO.
They were chosen in such a way as to span the gamut from dominated by a single reference configuration to pathological nondynamical correlation. Because of 
their small size, consideration of fairly large basis sets is possible. Results can be found in 
Table~\ref{tab:T4T5convergence}.

Let us first consider quasiperturbative connected quadruples, which we were able to obtain with cc-pVQZ basis sets for all systems. Considerable basis set sensitivity was seen, especially for systems with significant nondynamical correlation. The data give us reason to believe cc-pVQZ is close to basis set convergence, but this is clearly 
not a practically feasible basis set
for larger systems.

Generally the (Q) contributions increase monotonically with the basis set. 
However, in very polar systems (H$_2$O and HF, with MgO being an extreme case) anomalous nonmonotonic convergence is observed. The problem appears to largely go away if aug-cc-pVnZ basis sets are used on O and F. However, this may not be something one can do on a routine basis.

As we noted before\cite{w3} for $\hat{T}_4$ overall, extrapolation from cc-pVDZ and cc-pVTZ basis sets works well in some cases, but does more harm than good in cases like H$_2$O, HF, and especially MgO.

Scaling the cc-pVDZ result --- analogous to the scaled CCSDTQ/cc-pVDZ$-$CCSDT/cc-pVDZ difference in W3 theory --- does less harm but is clearly a rather crude approximation. However, there is a very high statistical correlation ($R^2=0.9945$) between the cc-pVTZ values and the numbers extrapolated from cc-pVTZ and cc-pVQZ basis sets. 

As expected, the CCSDTQ$-$CCSDT(Q) difference is quite small in systems dominated by a single reference configuration (like HF and H$_2$O). However, it becomes rather significant in systems with moderate to strong nondynamical correlation. This casts some doubt on the general applicability of CCSDT(Q) on its own as a post-CCSDT correction.

Fortunately, basis set convergence of the CCSDTQ$-$CCSDT(Q) difference is much more rapid than for the (Q) contribution, and even cc-pVDZ generally appears to be sufficient. This naturally suggests a simple additivity approximation:
\begin{eqnarray}
\Delta E(\hat{T}_4)&\approx& c_1 (E[\hbox{CCSDT(Q)/cc-pVTZ}] - E[\hbox{CCSDT/cc-pVTZ}] +\nonumber\\
&&+ E[\hbox{CCSDTQ/cc-pVDZ}] - E[\hbox{CCSDT(Q)/cc-pVDZ}])\label{eq:W4T4}
\end{eqnarray}
We found that this approximation works very well when fitted against the extrapolated basis set limit data. Eliminating the anomalous cases of BeO and MgO, we find $c_1=1.10$ to be optimal with $R^2=0.997$.
We also found that, in practice, all the steps in this additivity approximation are computationally feasible if  CCSDTQ/cc-pVDZ is feasible --- that is, any system tractable at the W3 level will be amenable to this improved connected quadruples term.

What about connected quintuple excitations? Results can be found in Table~\ref{tab:T5conv}.
We find that the CCSDTQ5$-$CCSDTQ difference converges quite rapidly with the basis set, and that even a simple double-zeta basis set comes quite close to the presumed basis set limit. In our experience, a CCSDTQ5/DZ calculation is feasible for any system that we were able to do CCSDTQ/cc-pVDZ for. 

How useful are quasiperturbative techniques here? We find the CCSDTQ5$-$CCSDTQ(5) difference to be essentially negligible in systems dominated by a single reference configuration. Even with fairly mild nondynamical correlation,
however, this difference behaves somewhat erratically.
If moderate to strong static correlation is present, however, we see exaggerated (5) contributions, with CCSDTQ5$-$CCSDTQ(5) being large in the opposite direction and $\hat{T}_5$ overall thus rather modest
(albeit significant enough for inclusion in anything purporting to improve on W3 theory).

Finally, the largest $\hat{T}_6$ contribution (Table~\ref{tab:T5conv}) is seen for singlet C$_2$ (about 0.06 kcal/mol). In less pathologically multireference systems, it can be neglected for all but the most accurate work (which would also require much greater accuracy in all other contributions). For systems with no or only mild static correlation, the quasiperturbative CCSDTQ5(6) method reproduces essentially the entire effect: in cases with severe nondynamical correlation, we note again that (6) tends to exaggerate the effect. 

\subsection{Connected quadruple excitations for the W3 set}

We now turn to the W3 set of molecules.
Connected quadruple excitation contributions to the total atomization energy 
can once again be found in Table~\ref{tab:T4T5convergence}. The largest contribution is found for ozone (3.81 kcal/mol) but it exceeds 2 kcal/mol for several other molecules, and 1 kcal/mol for surprisingly many. It was conjectured by Bak et al.\cite{Bak2000} and confirmed repeatedly before (e.g.\cite{w3,Ruden2003,Ruden2004,Crawford2006}) that the main reason for the good performance of CCSD(T) for many systems is error compensation between neglect of higher-order $\hat{T}_3$ effects (which generally decrease the atomization energy) and complete neglect of $\hat{T}_4$ (which systematically increases atomization energies). This is once more confirmed here.

Differences between the W3 estimate for $\hat{T}_4$ effects --- that is, $1.25(E[\hbox{CCSDTQ/cc-pVDZ}]-E[\hbox{CCSDT/cc-pVDZ}])$ --- and Eq.(\ref{eq:W4T4}) may reach 0.5 kcal/mol and are somewhat erratic in character. The W3 estimate was a stop-gap measure at the time for want of a better yet sufficiently cost-effective alternative. Eq.\ref{eq:W4T4} appears to have made it obsolete.

\subsection{Connected quintuple and higher excitations}

Connected quintuple excitation contributions to the total atomization energy for the W3 set of molecules can be found in Table~\ref{tab:T5conv}. The largest contribution found is for ozone, 0.41 kcal/mol. With one exception (OCS, decrease by 0.01 kcal/mol), $\hat{T}_5$ systematically increases the binding energy.

We were unable to obtain iterative $\hat{T}_6$ corrections for many of the 
systems. In light of what was noted for the diatomics above, and in light of the quasiperturbative $\hat{T}_6$ corrections which we were able to obtain
for most systems, $\hat{T}_6$ contributions are expected to be negligible compared to other intrinsic errors in our calculations.

\subsection{Diagonal Born-Oppenheimer corrections}

Valeev and Sherrill\cite{ValeevSherrill} studied correlation effects on diagonal Born-Oppenheimer corrections in some detail, and concluded that they should be 
very small for relative energies, and that the DBOCs should be adequately reproduced at the
Hartree-Fock level with a basis set of at least AVTZ quality. This is the level of theory considered by us. Contributions can be found in Table~\ref{tab:DBOC}.
(ROHF/AVDZ values are also given there, in order to show that the ROHF/AVTZ values are converged with respect to the basis set.)

 As expected, they are most important for the hydrides, and can reach or exceed 0.1 kcal/mol if multiple hydrogens are present. For the benzene molecule (not listed in Table \ref{tab:DBOC}), we computed a contribution of 0.24 kcal/mol. For semirigid closed-shell molecules, DBOC appears to systematically increase the binding energy, but decreases are seen for some open-shell and less rigid molecules.

\subsection{Larger basis sets for valence higher-order $T_3$ contributions}

In W3 theory, the CCSDT-CCSD(T) difference was extrapolated from cc-pVDZ and cc-pVTZ basis sets. In this work, we considered whether it would be worthwhile to extrapolate this contribution from larger cc-pVTZ and cc-pVQZ basis sets (at the expense of approximately a factor 32 in computational effort). As can be seen in Table~\ref{tab:T3conv}, the effect is in the 0.01 kcal/mol range in most cases, with B$_2$ ($-$0.11 kcal/mol),  
PH$_3$ (0.08 kcal/mol), and O$_2$ (0.07 kcal/mol) 
being the exceptions that prove the rule. We have therefore elected to retain cc-pVDZ and cc-pVTZ basis sets for this contribution.

\subsection{Post-CCSD(T) contributions to the core-valence contribution}

It was shown in the original W1 paper\cite{w1} that connected triple excitations are quite important for the core-valence contribution, accounting for as much as half the total. It can then not a priori be ruled out that post-CCSD(T) contributions to the inner-shell correlation energy would be non-negligible. Limited anecdoctal evidence gathered in the W3 paper\cite{w3} 
suggests possible contributions on the order of 0.05--0.10 kcal/mol, which might well be relevant for our purposes. Clearly, this point bears further scrutiny.

In addition, CCSDT-CCSD(T) differential core-valence contributions may resolve an ambiguity resulting from different CCSD(T) definitions. (See Appendix 1 for a more detailed discussion.)
Briefly, closed-shell CCSD(T) and CCSDT are uniquely defined, as are UCCSD(T) and UCCSDT, regardless of whether electrons are frozen. By contrast,  ROCCSD(T) is only uniquely defined if all electrons are correlated. If some are `frozen', two ambiguities arise: (1) whether standard or semicanonical orbitals are used; (2) if the latter, whether the `frozen core' orbitals are included or excluded from the semicanonicalization. In practice, performing ROCCSD(T) in a basis of standard orbitals requires an additional $O(N^7)$ step, so therefore the two main practical implementations (Watts-Gauss-Bartlett\cite{Wat93}, a.k.a. ACES II, and Werner-Hampel-Knowles\cite{Ham2000}, a.k.a. MOLPRO) use semicanonicalization. Their mutual nonequivalency arises on point (2), in that ACES II includes the frozen core orbitals in the semicanonicalization, while MOLPRO excludes them. 

ROCCSDT with all electrons correlated would be devoid of such ambiguities. However, basis set limit calculations at this level are manifestly not a practical approach.  Rather, we shall apply the following additivity approximation:

\begin{eqnarray}
E[\hbox{CCSDT}]_{\rm all}^{\rm limit}&\approx& E[\hbox{CCSD(T)}_{\rm val}^{\rm limit1}]+E[\hbox{CCSDT}-\hbox{CCSD(T)}]_{\rm val}^{\rm limit2}\nonumber\\&+&E[\hbox{CCSD(T)}_{\rm all}-\hbox{CCSD(T)}_{\rm val}]^{\rm limit3}\nonumber\\&+&E[\hbox{CCSDT}_{\rm all}-\hbox{CCSD(T)}_{\rm all}-\hbox{CCSDT}_{\rm val}+\hbox{CCSD(T)}_{\rm val}]^{\rm limit4}
\end{eqnarray}

The first term is the sum of our SCF, CCSD, and (T) limits. The second term is our valence $T_3-$(T) correction. The third term is our calculated core-valence correction, while the fourth and final term is the differential $T_3-$(T) contribution to the core-valence correction. The different `limit' labels reflect the fact that each component is obtained at a different {\em approximate} basis set limit --- if the true basis set limit were available in each case the `is approximately equal to' sign would change to an equality.

The all-electron CCSDT calculations involved in the final term strain our computational resources to the limit even for the heavier 1st-row systems: they will be effectively impossible for most 2nd-row systems. 

In Table~\ref{tab:coreT3T4}, we have gathered values for this final term using ACES II and MOLPRO definitions. It can be seen there that the corrections are generally negative for ACES and positive for MOLPRO, and that their absolute values are quite small even for significantly multireference systems. Only for pathologically multireference molecules like C$_2$ and BN do they even exceed 0.1 kcal/mol.  

The same Table also contains quasiperturbative $T_4$ differential contributions defined as follows:

\begin{eqnarray}
\Delta Q &\approx& E[\hbox{UCCSDT(Q)}-\hbox{UCCSDT}]_{\rm all}-E[\hbox{UCCSDT(Q)}-\hbox{UCCSDT}]_{\rm val}\nonumber\\
\Delta\Delta Q&=& \sum_{\rm atoms}\Delta Q_{\rm atom} - \Delta Q_{\rm molecule}
\end{eqnarray}

Differential $T_4$ contributions become somewhat significant for pathologically multireference molecules but are insignificant for our purposes otherwise. 

In summary, the post-CCSD(T) contribution to the core-valence correlation energy --- at least for first-row systems --- 
appears to be negligibly small in most cases. Its neglect will slightly bias TAEs upward if the ACES definition of valence CCSD(T) is used, and slightly downward if the MOLPRO definition is used instead. We have therefore elected to take an average. Specifically, we shall be using the MOLPRO definition throughout, except for adding in one-half the ACES$-$MOLPRO difference extrapolated from cc-pVDZ and cc-pVTZ basis sets. As can be seen in Table \ref{tab:ACESvsMOLPRO}, this difference 
converges extremely rapidly with the basis set, and values extrapolated from cc-pV\{D,T\}Z and cc-pV\{T,Q\}Z basis sets
are basically indistinguishable.

\subsection{Definitions of W4 theory and variants}

For consistency, as long as we are investing the quite formidable computational effort required for the improved post-CCSDT corrections, we might want to improve the SCF, CCSD(T), and core correlation contributions relative to those used in W2 and W3 theory.

The use of larger basis sets for the SCF and valence CCSD(T) correlation was already considered in some detail in the W3 paper\cite{w3}. With the hardware detailed in the Methods section, we found that we could carry out CCSD/AV6Z and CCSD(T)/AV5Z calculations using conventional algorithms for all the systems considered in the present paper.

We thus propose the following protocol for W4 theory:
\begin{itemize}
\setlength{\itemsep}{0pt}
\item the ROHF-SCF contribution is extrapolated from AV5Z and AV6Z basis sets using the recently proposed Martin-Karton modification\cite{MKtca} of Jensen's extrapolation formula\cite{JensenExtrap}:
\begin{equation}
E_{{\rm HF},L}=E_{{\rm HF},\infty}+A(L+1)\exp(-9\sqrt{L})
\end{equation}
\item the RCCSD valence correlation energy is calculated using AV5Z and AV6Z basis set, using the Watts-Gauss-Bartlett definition\cite{Wat93} for open-shell systems. 
Following the suggestion of Klopper\cite{Klopper}, it is partitioned in singlet-coupled pair energies, triplet-coupled pair energies, and $\hat{T}_1$ terms. (The term linear in the single excitations $\hat{T}_1$ in the CCSD
equations is nonzero for open-shell CCSD calculations using semicanonical
orbitals, see, e.g., Ref.\cite{T1energy}.) The $\hat{T}_1$ term (which exhibits very weak basis set dependence) 
is simply set equal to that in the largest basis set, while the singlet-coupled and triplet-coupled pair energies are extrapolated by the expression
\begin{equation}
E_\infty=E(L)+\frac{E(L)-E(L-1)}{(L/L-1)^\alpha-1}\label{eq:Lextrap}
\end{equation}
with $\alpha_S$=3 and $\alpha_T$=5, and $L$ set equal to the maximum angular momentum present in each basis set (i.e., 5 for AV5Z and 6 for AV6Z). These expressions are physically motivated by the partial-wave expansion of pair correlation energies in helium-like atoms\cite{Sch63,Hil85,Kut92} as well as by empirical observation\cite{Hal98,w1}.
\item the (T) valence correlation energy was extrapolated using the same expression with $\alpha$=3, from AVQZ and AV5Z calculations. Note that only a CCSD calculation is required in the largest (AV6Z) basis set. For open-shell systems, the Werner-Knowles-Hampel (a.k.a. MOLPRO) definition\cite{Ham2000} of the restricted open-shell CCSD(T) energy is employed throughout, rather than the original Watts-Gauss-Bartlett\cite{Wat93} (a.k.a. ACES II) definition, unless indicated otherwise;
\item the CCSDT$-$CCSD(T) difference is extrapolated using Eq.(\ref{eq:Lextrap})
from CCSDT$-$CCSD(T) differences with PVDZ and PVTZ basis sets
\item the $\hat{T}_4$ difference was estimated from eq.(\ref{eq:W4T4}). For this contribution, UHF reference determinants are used for want of an restricted open-shell CCSDT(Q) code. In any case, RCCSDT and UCCSDT energies are very close for all the systems considered here, and we have no reason to believe that there is any significant error introduced in this term by spin contamination.
One might argue about the use of RCCSDTQ/cc-pVDZ versus UCCSDTQ/cc-pVDZ in Eq.(1). We have considered both options. 
For the closed-shell molecules, the results are equivalent to within less than 0.01 kcal/mol, but one can see larger differences for radicals with significant spin contamination. For radicals like O$_2$ and NO$_2$, the ROCCSDTQ based estimate yields results substantially closer to the very precisely known experimental values, and we have therefore retained this choice.
\item the $\hat{T}_5$ contribution was estimated from CCSDTQ5/DZ calculations
\item the difference between ACES II and MOLPRO definitions of the valence ROCCSD(T) definition is computed at the CCSD(T)/cc-pVTZ and CCSD(T)/cc-pVQZ levels and extrapolated using Eq.(\ref{eq:Lextrap}) (with $\alpha$=3). One-half of this contribution is added to the final result.
\item the inner-shell correlation contribution was extrapolated using Eq.(\ref{eq:Lextrap}) (with $\alpha$=3) from CCSD(T)/aug-cc-pwCVTZ and CCSD(T)/aug-cc-pwCVQZ calculations
\item the scalar relativistic contribution is obtained from the difference between nonrelativistic CCSD(T)/aug-cc-pV(Q+d)Z and second-order Douglas-Kroll
CCSD(T)/DK-aug-cc-pV(Q+d)Z calculations. We found that there is essentially no difference between such contributions obtained using the unpublished Martin-Oren Douglas-Kroll optimized basis sets\cite{oren-martin} and the publicly available PNNL Douglas-Kroll contracted correlation consistent basis sets\cite{Dyall}, and have chosen the latter as the nonrelativistic result can be recycled from the (T) step.
\item atomic spin-orbit coupling terms are taken from the experimental fine structure. The spin-orbit splitting constant for NO --- the only molecule
considered here with first-order spin-orbit coupling --- was taken from Ref.\cite{Hub79}. It was previously shown\cite{w1} that both atomic and molecular spin-orbit coupling constants can be computed as well without significant loss of accuracy.
\item Finally, diagonal Born-Oppenheimer corrections are obtained from 
ROHF/AVTZ calculations.
\end{itemize}

In the HEAT paper (and `superHEATed' methods discussed there\cite{HEAT}), 
unrestricted reference wave functions are used throughout and inner-shell electrons are only frozen in the post-CCSD(T) correlation treatments. Only
first-row molecules were considered there.

Finally, we note that for electron affinities --- as well as for systems 
like MgO that are both very polar and pathologically multireference in
character --- the use of augmented basis sets in the $T_3-(T)$ and $T_4$ 
steps is very strongly recommended. It was previously shown for Na, Mg, and 
especially K and Ca systems that the inner-valence shell of these metals should
be added to the valence shell\cite{Radom,IOM}.

\subsection{Validation against Active Thermochemical Tables data}

Individual components of the final W4 results can be found in Table~\ref{tab:W4components}. A comparison between the final W4 values and the Active Thermochemical Tables data (with associated 95\% confidence intervals) is given in 
Table~\ref{tab:W4results}.\cite{H2footnote}

A remark is due on the anharmonic zero-point vibrational energies used. As the uncertainty in some of the ZPVEs employed in the W1/W2/W3 series of papers is on the order of 0.1 kcal/mol, we have re-examined the data available: source details are given in the footnotes to Table~\ref{tab:W4results}. For two species, CH$_3$ and CH$_4$, the best available data represent revisions in excess of 0.1 kcal/mol.

For the first-row systems, agreement between W4 and ATcT data can only be described as excellent, with an RMSD (root mean square deviation) of 
0.083 kcal/mol
and a mean absolute deviation (MAD) of 
0.065 kcal/mol.
(This implies a 95\% confidence interval of about 0.16 kcal/mol.) The mean signed deviation (MSD) is essentially zero, at 
$-0.011$ kcal/mol, 
suggesting that the method is free of systematic bias.

If we add in the 2nd-row species, RMSD and MAD go up to 0.15 and 0.09 kcal/mol, respectively. However, basically all the extra error is caused by a single molecule, ClCN, for which the ATcT value is mostly based on prior theoretical calculations. If we remove this data point, RMSD and MAD drop to values equivalent to those for the first row alone.\cite{SOandSO2}

Including the remaining W3 species, and using the earlier experimental data compiled in Ref.\cite{w1} for species lacking an ATcT value, we find both RMSD and MAD to be 0.15 kcal/mol with ClCN included, both dropping to 0.10 kcal/mol with ClCN excluded. 

Overall, excepting ClCN, the discrepancy between W4 and experiment reaches or exceeds 1 kJ/mol for only 2 species: PH$_3$ (+0.31 kcal/mol) and O$_3$ (-0.24 kcal/mol).
For PH$_3$, the calculated result is still within the experimental error bar ($\pm$0.41 kcal/mol). 

For the first-row and part of the second-row species, we can add in the CCSDT/cc-pCVTZ$-$CCSD(T)/cc-pCVTZ differential core-valence contribution to the total atomization energy, thus obtaining what we term W4.2 theory. (For second-row elements, the cc-pwCVTZ basis set was used.) Here, differences between ACES and MOLPRO definitions of 
CCSD(T) are on the order of 0.01 kcal/mol or less, as they ought to be. For the ATcT 1st-row systems, W4.2 theory represents a slight improvement over W4 theory: the W4.2 error statistics are 
RMSD=0.067, MSD=+0.003, MAD=0.056 kcal/mol. 
(This implies a 95\% confidence interval of about 0.13 kcal/mol.) As expected, the largest improvement is seen for ozone (from 0.24 to 0.14 kcal/mol): we expect that this error would be reduced further if a differential $\hat{T}_4$ contribution to the core correlation could be included and/or a  valence $\hat{T}_6$ contribution could be added. The former requires about 92 billion determinants in the cc-pCVTZ basis set, which is sadly beyond our available computational resources.
(An at best semiquantitative estimate can be obtained at the CCSDT(Q)/cc-pCVDZ level: this suggests a narrowing of the gap by 0.04 kcal/mol, to a rather pleasing 0.1 kcal/mol.) As for the latter, a crude estimate of the $\hat{T}_6$ contribution can be obtained as one-fourth the $\hat{T}_5$ contribution (see Section~\ref{sec:diag} below), i.e., a further increase by 0.1 kcal/mol which would bridge the gap essentially completely.

At the W4.2 level, our biggest remaining error is for Cl$_2$ (-0.17 kcal/mol). Using larger basis set CCSDT$-$CCSD(T) and $T_4$ corrections reduces the discrepancy to -0.12 kcal/mol, which is further reduced to -0.10 kcal/mol when a larger basis set is used for $T_5$ as well. 

Based on this observation, we are quoting an additional set of results, labeled `W4.3' in the Table. These represent the results of more rigorous accounting for valence post-CCSD(T) effects --- all other contributions are the same as for W4.2. Specifically, we are (a) using our `best estimate' valence $\hat{T}_4$ contributions; (b) CCSDT-CCSD(T) is extrapolated from cc-pV\{T,Q\}Z basis sets; (c) the $\hat{T}_5$ contribution is obtained with the cc-pVDZ basis set; (d) an approximate $\hat{T}_6$ contribution was obtained at the CCSDTQ5(6)/cc-pVDZ level. 

Relative to W4.2, the performance of W4.3 is somewhat mixed. For some systems --- notably O$_2$, Cl$_2$, ClF, and to a lesser extent, CO --- agreement with experiment is markedly improved, for others (notably N$_2$) it deteriorates.  For the remainder of the systems, the W4.2--W4.3 difference is either too small to affect anything, or the two values err on opposite sides of experiment. In all probability, for a W4.3 type method to yield any further improvement, the valence CCSD extrapolations would need to be rendered still more rigorous, as we believe the W4.2--W4.3 differences are on the order of the valence CCSD uncertainty in a number of our systems. 

Finally, for some systems we are actually able to add the (Q)/cc-pCVTZ contribution to the 
core-valence correlation term. We have not tabulated these results, as they do not
appreciably affect the systems given in the Table for which we could obtain them.
Their inclusion might be worthwhile in systems like C$_2$ or BN, or --- as noted above, and if the daunting
computational requirements could be met --- ozone. 

\subsection{An improved and more cost-effective W3 theory}

Stanton and coworkers have argued\cite{Bomble2005} that CCSDT(Q) might be closer to FCI than CCSDTQ, for similar reasons as in the case of CCSD(T) vs. CCSDT\cite{Stanton1997}. This begs the question if one could not derive a lower-cost variant of W3 theory involving CCSDT(Q)/cc-pVDZ. 

Eliminating the anomalous cases of BeO and MgO, we find that $1.025(E[{\rm CCSDT(T)/cc-pVDZ}]-E[{\rm CCSDT/cc-pVDZ}])$ 
reproduces the W4 estimate of post-CCSDT correlation effects with $R^2=0.974$, the largest deviations being seen for the B$_2$ and CS molecules. When these latter two systems are eliminated, the coefficient changes to 1.012 (only semantically different from unity for our purposes), with $R^2=0.991$. Carrying out regression instead to our `best estimate' post-CCSDT contributions, we find three clear outliers (B$_2$, CS, and Cl$_2$): upon eliminating them, we find a coefficient of 1.000 with $R^2$=0.990. We thus essentially recover the HEAT\cite{HEAT} post-CCSDT contribution.

The component breakdown for W3.2 theory is given in the Supporting Information: the final W3.2 results can be found in Table~\ref{tab:W4results}, compared with W4 and experiment. Occasionally a substantial deviation from W4 theory is seen (notably for B$_2$, somewhat less so for CS) but by and large, W3.2 theory appears to be closer to W4 theory than the original W3 theory. 

We might then define W2.2 theory as W3.2 theory without the post-CCSD(T) contributions.

For the ATcT species minus ClCN, W3.2 theory obtains an RMSD=0.16 kcal/mol and MAD=0.12 kcal/mol.
This is 
competitive with W3 theory itself, at considerably reduced computational cost. Neglecting post-CCSD(T) correlation effects altogether, we obtain W2.2 theory. The latter does very well, as expected, for species dominated by a single reference determinant, but yields unacceptable errors for species like N$_2$O,
NO$_2$, and especially ozone. 

Finally, let us consider W4 theory in which just the post-CCSDT contribution is approximated as unscaled (Q)/cc-pVDZ; ``W4lite'', as it were. Its performance is
intermediate between W3.2 and full W4: for the ATcT species less ClCN, MAD=0.09 and RMSD=0.12 kcal/mol. The performance differential with W3 is primarily due to the larger basis sets employed in the various CCSD(T) level contributions. ``W4lite'' may be applicable to some systems where CCSDT(Q)/cc-pVTZ and/or CCSDTQ/cc-pVDZ are prohibitive in computational cost.

A proposed convergent hierarchy of methods (at increasing cost) might thus be W1 $\rightarrow$ W2.2 $\rightarrow$ W3.2 $\rightarrow$ W4lite $\rightarrow$ W4.

\subsection{A simple energy-based diagnostic for nondynamical correlation effects\label{sec:diag}}

A number of diagnostics have been proposed for nondynamical correlation character of a system, such as the ${\cal T}_1$
diagnostic of Lee and Taylor\cite{Lee89t1} and the ${\cal D}_1$ and ${\cal D}_2$ diagnostics of Nielsen and Janssen\cite{D1diagnostic}. Other researchers (particularly those involved with multireference methods) consider such indicators
as the largest $T_2$ amplitudes or the HOMO and LUMO natural orbital occupations. Such diagnostics are gathered in Table~\ref{tab:diagnostics}, together with some energy-based quantities --- specifically, the percentages of the total atomization energy accounted for by SCF, (T) triples, and $\hat{T}_4+\hat{T}_5$. 

The latter is probably the best yardstick for imperfections in CCSD(T) for our purposes, but of course it is an {\em a posteriori} criterion. An {\em a priori} indicator for whether the formidable computational effort entailed by post-CCSD(T) methods is necessary would be highly desirable. 

Of the various diagnostics displayed in Table~\ref{tab:diagnostics}, the \%TAE[(T)] criterion appears to be the best predictor for \%TAE[$T_4+T_5$], with a squared correlation coefficient $R^2=0.791$. If the somewhat anomalous systems BeO and MgO are deleted, we find a linear regression \%TAE[$T_4+T_5$]$\approx$0.126\%TAE[(T)],  with the squared correlation coefficient increasing to a respectable $R^2$=0.941. (The regression slope including BeO and MgO is 0.11.) 

The second-best predictor appears to be \%TAE[SCF]. With BeO and MgO eliminated, we find a linear regression \%TAE[$T_4+T_5$] $\approx -0.0199$\%TAE[SCF] + 1.5997, with $R^2=0.810$. 

The following approximate predictors for specific higher-order excitations can be obtained in the same manner: 
\%TAE[$T_5$] $\approx$ 0.095 \%TAE[$T_4$]  (eliminating just OCS as an outlier, $R^2$=0.922) and \%TAE[$T_6$] $\approx$ 0.24 \%TAE[$T_5$] (no outliers, $R^2$=0.917). While these are obviously no replacement for explicit calculations, they are quite useful in establishing whether a post-CCSD(T) approach such as W3 or W4 theory is required in the first place.

Note also that, as an error estimate for W2 theory, 0.126\%TAE[(T)] is rather pessimistic, as \%TAE[$T_4+T_5$] will 
be compensated to greater or lesser extent by higher-order connected triple excitations.

Turning now to the other diagnostics, we note clear failures for all of them. We note, for instance, the deceptively low
${\cal T}_1$ diagnostic of $F_2$, or the fact that the largest $T_2$ amplitude for H$_2$CO is larger than that for O$_2$, despite the latter having almost an order of magnitude greater \%TAE[$T_4+T_5$]. The LUMO natural orbital occupation
appears to track \%TAE[$T_4+T_5$] somewhat more consistently, although it requires the CCSD response to calculate
(which is not available in the two most commonly used quantum chemical codes, Gaussian 03 and MOLPRO).

We conclude that the simple energy-based diagnostics \%TAE[(T)] and \%TAE[SCF] --- which require no additional software or computational effort to calculate --- are the most useful, at least for thermochemical applications.

For general `user convenience', let us address how to interpret the numerical values of \%TAE[SCF]
and \%TAE[(T)] in a qualitative sense. The data in Table \ref{tab:diagnostics}
suggest that:
\begin{itemize}
\setlength{\itemsep}{0pt}
\item \%TAE[(T)] below 2\% indicate systems dominated by dynamical correlation
\item \%TAE[(T)] between 2\% and about 4--5\% indicate mild nondynamical correlation
\item \%TAE[(T)] between 4--5\% and about 10\% indicate moderate,
\item and values in excess of 10\% severe, nondynamical correlation
\end{itemize}
The \%TAE[SCF] data offer less detail, but as a rule of thumb, \%TAE[SCF] above 66.7\% (two-thirds) suggest systems largely or wholly dominated by dynamical
correlation, and \%TAE[SCF] below 20\% (one-fifth) --- particularly 
negative values --- indicate severe nondynamical correlation.

Finally, in light of the fairly good correlation (without outliers) between
$\hat{T}_6$ and $\hat{T}_5$ and the small numbers involved, 
we could consider adding 0.24$\Delta$TAE$(\hat{T}_5)$ as an empirical 
correction for connected sextuple excitations. Doing so somewhat violates the
general spirit of the `W$n$ theory' family (as we are estimating a contribution to the Hamiltonian rather than explicitly calculating it), and we shall not report any such values (the interested reader can easily obtain them using a pocket calculator and Tables \ref{tab:T5conv} and \ref{tab:W4results}). 
For such a putative `W4as theory' (`as' for `approximate sextuples'), error statistics 
for the 1st-row ATcT systems would indeed be improved somewhat compared to regular W4: MSD=-0.008, MAD=0.059, RMSD=0.069 kcal/mol. (Also including
the 2nd-row ATcT species, we have MSD=-0.005, MAD=0.059, RMSD=0.071 kcal/mol.) For a putative `W4.2as theory', the corresponding 1st-row error statistics are MSD=-0.016, MAD=0.049, RMSD=0.060 kcal/mol. As noted above, the single most
notable improvement would be for ozone.

\subsection{Prospects for application to heavier-element systems}

Moving beyond Ar (or perhaps Ca) in the Periodic Table  without sacrificing accuracy will entail some additional considerations.

One of them is second-order spin-orbit coupling. It has been neglected for the systems considered here: for HF, HCl, F$_2$, and Cl$_2$, however, we can compare the sum of scalar relativistic corrections and atomic spin-orbit coupling to relativistic corrections obtained by Visscher et al.\cite{vis1,vis2} from full four-component relativistic coupled cluster calculations. In all cases, differences are well below 0.1 kcal/mol. However, previous work at PNNL\cite{Dixon2003} on bromine and iodine compounds suggested approximate second-order spin-orbit contributions for Br$_2$ and I$_2$ of 0.4 and 2.0 kcal/mol, respectively, which clearly cannot be neglected with impunity.

Secondly, for such heavy elements, basis set superposition error may need 
to be accounted for in extrapolations, although small-core ECP-based
basis sets may offer a better alternative. 

Thirdly, more sophisticated basis set extrapolation schemes may be called for, especially in transition metal systems. Research on this issue is currently
under way in our laboratory.

Fourthly, for more heavily multireference transition metal systems, 
CCSD(T) may no longer be adequate for geometry optimizations.

\section{Conclusions}

We have proposed a new computational thermochemistry protocol, named W4 theory, for first- and second-row main group elements and validated it against benchmark values obtained from the 
latest version of the ATcT (Active Thermochemical Tables) network. For 
key species with well-established atomization energies, W4 theory reaches an average accuracy better than 0.1 kcal/mol, including for such difficult
species as ozone. Most systems for which the earlier W3 theory is feasible are amenable to W4 calculations.

The recent CCSDT(Q) method offers a cost-effective way of estimating $\hat{T}_4$, but is insufficient by itself if the molecule exhibits some nondynamical
correlation. The latter considerably slows down basis set convergence for $\hat{T}_4$, and anomalous basis set convergence in highly polar systems makes 
two-point extrapolation procedures unusable. However, we found that the CCSDTQ$-$CCSDT(Q) difference converges quite rapidly with the basis set, and that the formula 1.10[CCSDT(Q)/cc-pVTZ+CCSDTQ/cc-pVDZ-CCSDT(Q)/cc-pVDZ] offers a very reliable as well as fairly cost-effective estimate of the basis set limit $\hat{T}_4$ contribution. 

The $\hat{T}_5$ contribution converges very rapidly with the basis set, and even a simple double-zeta basis set appears to be adequate. The largest $\hat{T}_5$ contribution found in the present work is on the order of 0.5 kcal/mol (for ozone). DBOC corrections are significant at the 0.1 kcal/mol level in hydride systems.

Post-CCSD(T) contributions to inner-shell correlation are quite small, except in
systems with severe nondynamical correlation.

We further propose low-cost versions, which we denote W4lite, W3.2, and W2.2 theory. 
W3.2 theory is found to be about as reliable as the original W3 theory (at much reduced cost), and supersedes the latter. W2.2 theory is essentially W2w theory\cite{w3} with a diagonal Born-Oppenheimer correction added. W4lite is intermediate in accuracy between W3.2 and full W4, and may be applicable to some systems beyond the reach of full W4.
The sequence W1$\rightarrow$W2.2$\rightarrow$W3.2$\rightarrow$W4lite$\rightarrow$W4 forms a convergent hierarchy of computational thermochemistry methods, where all steps required in W2.2 theory can be recycled for W3.2 theory, and all steps required in the latter can be recycled for W4 theory.

Finally, we have considered various diagnostics for the importance of post-CCSD(T) correlation effects, and find the simple energy based criteria \%TAE[(T)] and \%TAE[SCF] to be the most useful for our purposes.

\acknowledgments

Research was supported by the Israel Science Foundation (grant 709/05), the Minerva Foundation (Munich, Germany), and the Helen and Martin Kimmel Center for Molecular Design. JMLM is the incumbent of the Baroness Thatcher Professorial Chair of Chemistry and a member {\em ad personam} of the Lise Meitner-Minerva Center for Computational Quantum Chemistry. The work at Argonne National Laboratory, together with the underlying fundamental thermochemical development of the ATcT approach, was supported by the U.S. Department of Energy, Division of Chemical Sciences, Geosciences, and Biosciences of the Office of Basic Energy Sciences, under Contract No. W-31-109-ENG-38. Development of inherent computer-science aspects of ATcT and the underpinning data management technologies were supported by the U.S. Department of Energy, Division of Mathematical, Information, and Computational Science of the Office of Advanced Scientific Computing Research, under Contract No. W-31-109-ENG-38 (Argonne) as part of the multi-institutional Collaboratory for Multi-Scale Chemical Science (CMCS), which is a project within the National Collaboratories Program of the U.S. Department of Energy.

The authors would like to thank
Prof. John F. Stanton (U. of Texas, Austin), Prof. Peter R. Taylor (Warwick U.), Prof. Donald G. Truhlar (U. of Minnesota),
and Prof. Hans-Joachim Werner (U. of Stuttgart) for helpful discussions and Dr. Mih\'aly Kallay (Budapest) for early access to a new version of MRCC and
kind assistance with the code. The general ATcT development has benefited from the support and effort of numerous past and present CMCS Team members.

The research presented in this paper is part of ongoing work
in the framework of a Task Group of the 
International Union 
of Pure and Applied Chemistry on
'Selected free radicals and critical intermediates: thermodynamic
properties from theory and experiment'
(2000-013-2-100, renewal 2003-024-1-100).
See Ref.\cite{iupac1} for further details.

\section*{Appendix I: Effect of different reference wave functions on energies obtained from high-order coupled cluster methods}

For open-shell species, differences between results obtained with different reference wave functions are liable to cause some confusion, which we would like to address here. Some illustrative data can be found in Table \ref{tab:TABLEV}.

At the full CI level with all electrons correlated, restricted and unrestricted wave functions should give the same answers, since the ROHF and UHF orbitals are related by a unitary transformation, under which FCI is invariant.

This will no longer be the case when core orbitals are frozen: even at the FCI level there will be small but finite differences between the total energies. For the first-row atoms B--F, we found these to be in the 20--50 microhartree range.
Small as these numbers may seem, for a polyatomic they may add up to a nontrivial discrepancy between ROFCI and UFCI atomization energies.

Likewise, when core orbitals are frozen, results --- even at the FCI level --- become slightly dependent on whether
one uses standard or semicanonical orbitals.

For fully iterative coupled cluster methods such as CCSD, CCSDT, CCSDTQ,\ldots, there will be nonzero differences between restricted open-shell and unrestricted energies due to spin contamination, even when all electrons are correlated. The difference tapers off to zero as the FCI limit is approached: for the first-row atoms B--F, the ROCCSDT $-$ UCCSDT differences with all electrons correlated are already down to the 100 nanohartree range. 

For CCSDT, CCSDTQ, \ldots calculations with frozen cores, the difference between restricted open-shell and unrestricted data will normally be dominated by the intrinsic ROFCI - UFCI difference, which even for CCSDT (a fortiori for CCSDTQ and CCSDTQ5) is several orders of magnitude larger than the spin contamination discrepancy.

Quasiperturbative methods entail an additional complication. For instance, in the CCSD(T) method, the underlying CCSD is invariant to unitary rotations within the orbital space for both restricted open-shell and unrestricted reference orbitals, but the (T) term is {\em not} invariant in the case of a restricted open-shell reference. Both of the leading implementations of ROCCSD(T) --- Watts et al.\cite{Wat93} as implemented in Aces II and other codes, and Hampel et al.\cite{Ham2000} as implemented in MOLPRO --- involve transformation from standard to semicanonical orbitals. The ROCCSD(T) energies of these two implementations are equivalent when all electrons are correlated. However, in the case of frozen core orbitals, the definitions are nonequivalent: while Aces II carries all orbitals in the semicanonicalization and then freezes out the {\em semicanonical} core orbitals in the integral transformation and coupled cluster steps, MOLPRO skips the core orbitals even at the semicanonicalization stage\cite{WernerPC}. The resulting CCSD(T) atomization energies are slightly different (Table \ref{tab:TABLEV}), up to about 0.1 kcal/mol. (The ROCCSD(T) and UCCSD(T) data are rather more different, as expected.)

What are the consequences of all this for a computational thermochemistry protocol like W4? First of all, the discrepancy between approximate ROFCI and UFCI limits in the valence-only steps will be offset by an (almost) identical but opposite discrepancy in the core-valence correction. (In practice, in a putative UW4 method a small fraction of the total atomization energy would move from the valence correlation to the inner-shell correlation contribution.)

Secondly, the inequivalence between the (T) implementations will introduce inequivalences at three stages: two nearly equal but opposite changes in the valence (T) and core-valence correlation steps, respectively, and another in the valence CCSDT$-$CCSD(T) difference. If we were including a core-valence CCSDT$-$CCSD(T) difference as well, we would once again have approximate cancellation of the inequivalences (as we do find in this work for W4.2 theory). In W4 theory, we are neglecting the latter term, so a slightly different result would be obtained depending on whether one used,
in the CCSDT$-$CCSD(T) step, CCSD(T) energies from ACES II or MOLPRO, or an average of the two (as in the present work).

\squeezetable
\begin{table}

\caption{Basis set convergence of CCSDTQ$-$CCSDT, CCSDT(Q)$-$CCSDT and CCSDTQ$-$CCSDT(Q)  ($\Delta$TAE, in kcal/mol)\label{tab:T4T5convergence}}
\begin{tabular}{l|cccc|cccc|ccc|cc}
\hline\hline
 & \multicolumn{4}{c}{CCSDTQ $-$ CCSDT} &  \multicolumn{4}{c}{CCSDT(Q) $-$ CCSDT} & \multicolumn{3}{c}{CCSDTQ $-$ CCSDT(Q)} & \multicolumn{2}{c}{Best $\hat{T}_4$ estimate} \\
 & \multicolumn{4}{c}{~} &  \multicolumn{4}{c}{~} & \multicolumn{3}{c}{~}  & \multicolumn{2}{c}{~~} \\
Basis set & DZ & PVDZ & PVTZ & PVDZ & DZ & PVDZ & PVTZ & PVQZ & DZ & PVDZ & PVTZ & (a) & Eq.(\ref{eq:W4T4})\\
Reference & UHF & UHF & UHF & ROHF & UHF & UHF & UHF & UHF & UHF & UHF & UHF &  & \\
\hline
H$_2$O & 0.28 & 0.24 & 0.17 & 0.24 & 0.29 & 0.26 & 0.19 & 0.21 & -0.01 & -0.02 & -0.02 & 0.21 & 0.18  \\
B$_2$ & 1.08 & 0.99 & 1.19 & 0.99 & 0.88 & 0.91 & 1.16 & 1.22 & 0.20 & 0.08 & 0.03 & 1.29 & 1.37  \\
C$_2$H$_2$ & 0.64 & 0.54 & 0.65 & 0.54 & 0.68 & 0.62 & 0.71 &  & -0.05 & -0.08 &  &  & 0.70  \\
CH$_3$ & 0.07 & 0.06 & 0.05 & 0.06 & 0.06 & 0.06 & 0.05 &  & 0.01 & 0.00 & 0.00 &  & 0.05  \\
CH$_4$ & 0.08 & 0.07 & 0.07 & 0.07 & 0.08 & 0.08 & 0.07 &  & 0.00 & 0.00 & 0.00 &  & 0.08  \\              
CH & 0.04 & 0.03 & 0.03 & 0.03 & 0.03 & 0.03 & 0.03 & 0.03 & 0.01 & 0.00 & 0.00 & 0.04 & 0.03  \\       
CO$_2$ & 1.11 & 0.99 &  & 0.99 & 1.27 & 1.21 & 1.17 &  & -0.16 & -0.22 &  &  & 1.04  \\
CO & 0.50 & 0.53 & 0.56 & 0.53 & 0.55 & 0.63 & 0.65 & 0.70 & -0.04 & -0.10 & -0.09 & 0.64 & 0.61  \\
F$_2$ & 1.00 & 0.82 & 0.80 & 0.82 & 1.06 & 0.93 & 0.91 & 0.98 & -0.06 & -0.11 & -0.11 & 0.92 & 0.89  \\    
HF & 0.16 & 0.17 & 0.09 & 0.17 & 0.17 & 0.19 & 0.11 & 0.12 & 0.00 & -0.02 & -0.02 & 0.11 & 0.10  \\
N$_2$ & 1.25 & 0.87 & 0.94 & 0.87 & 1.44 & 1.03 & 1.09 & 1.16 & -0.19 & -0.16 & -0.15 & 1.06 & 1.03  \\
NH$_3$ & 0.22 & 0.17 & 0.16 & 0.17 & 0.23 & 0.19 & 0.16 & 0.19 & -0.01 & -0.02 & -0.01 & 0.19 & 0.16  \\
N$_2$O & 2.16 & 1.75 &  & 1.75 & 2.62 & 2.20 & 2.26 &  & -0.46 & -0.46 &  &  & 1.98  \\
NO & 0.86 & 0.72 & 0.75 & 0.75 & 0.97 & 0.88 & 0.91 & 0.98 & -0.11 & -0.15 & -0.16 & 0.87 & 0.84  \\    
O$_2$ & 1.16 & 1.00 & 0.96 & 1.08 & 1.26 & 1.12 & 1.09 & 1.16 & -0.10 & -0.13 & -0.13 & 1.08 & 1.06  \\    
O$_3$ & 3.38 & 3.21 &  & 3.21 & 4.26 & 4.13 & 4.38 &  & -0.88 & -0.92 &  &  & 3.81  \\
C$_2$ & 0.95 & 1.59 & 2.12 & 1.59 & 2.29 & 2.66 & 3.22 & 3.35 & -1.34 & -1.07 & -1.10 & 2.35 & 2.37  \\
BN & 0.72 & 1.38 & 1.87 & 1.38 & 1.53 & 2.46 & 3.03 & 3.17 & -0.81 & -1.09 & -1.16 & 2.13 & 2.13  \\    
MgO & 1.31 & 1.57 & 1.74 & 1.57 & 2.45 & 2.85 & 2.85 & 2.76 & -1.14 & -1.28 & -1.11 & 1.58 & 1.72  \\
BeO & 0.53 & 0.69 & 0.67 & 0.69 & 1.11 & 1.38 & 1.16 & 1.13 & -0.59 & -0.69 & -0.49 & 0.61 & 0.52  \\
CN & 1.22 & 0.86 & 1.00 & 0.84 & 1.24 & 1.24 & 1.44 & 1.52 & -0.03 & -0.38 & -0.44 & 1.13 & 1.16  \\
NO$_2$ & 1.79 & 1.66 &  & 1.72 & 2.14 & 2.05 & 2.12 &  & -0.35 & -0.39 &  &  & 1.90  \\
Cl$_2$ & 0.11 & 0.24 & 0.41 & 0.24 & 0.10 & 0.26 & 0.43 & 0.49 & 0.00 & -0.02 & -0.02 & 0.51 & 0.44  \\
ClF & 0.35 & 0.39 & 0.42 & 0.39 & 0.37 & 0.44 & 0.47 & 0.51 & -0.02 & -0.05 &  &  & 0.46  \\
CS & 0.13 & 0.50 & 0.87 & 0.50 & 0.09 & 0.59 & 0.98 & 1.08 & 0.04 & -0.09 & -0.10 & 1.05 & 0.98  \\     
H$_2$S & 0.03 & 0.08 & 0.14 & 0.08 & 0.03 & 0.08 & 0.13 & 0.15 & 0.00 & 0.00 & 0.00 &  & 0.15  \\              
HCl & 0.02 & 0.07 & 0.09 & 0.06 & 0.02 & 0.07 & 0.09 & 0.10 & 0.00 & 0.00 & 0.00 & 0.11 & 0.10  \\
HOCl & 0.49 & 0.48 &      & 0.48 & 0.52 & 0.54 & 0.59 &  & -0.03 & -0.06 &  &  & 0.58  \\
PH$_3$ & 0.03 & 0.05 & 0.09 & 0.05 & 0.02 & 0.05 & 0.09 &  & 0.00 & 0.00 & 0.00 &  & 0.09  \\
SO & 0.90 & 0.68 & 0.75 & 0.73 & 0.95 & 0.80 & 0.88 & 0.94 & -0.05 & -0.12 &  & 0.86 & 0.84  \\
SO$_2$ & 1.68 & 1.44 &  & 1.44 & 2.15 & 1.81 & 1.79 &  & -0.47 & -0.37 &  &  & 1.56  \\
OCS & 0.85 & 0.98 &  & 0.98 & 0.95 & 1.16 & 1.39 &  & -0.10 & -0.18 &  &  & 1.32  \\
ClCN & 1.09 & 0.94 &  & 0.94 & 1.26 & 1.14 & 1.33 &  & -0.17 & -0.20 &  &  & 1.24  \\
C$_2$H$_4$ & 0.27 & 0.33 &  & 0.33 & 0.29 & 0.37 & 0.42 &  & -0.02 & -0.04 &  &  & 0.43  \\
H$_2$CO & 0.56 & 0.50 &  & 0.50 & 0.62 & 0.60 & 0.59 &  & -0.06 & -0.09 &  &  & 0.54  \\
HNO & 0.99 & 0.86 &  & 0.85 & 0.62 & 0.99 & 1.03 & 1.10 & -0.10 & -0.14 &  &  & 0.99  \\
\hline\hline
\end{tabular}
\begin{flushleft}
(a) Extrapolated from CCSDT(Q)/cc-pVTZ and CCSDT(Q)/cc-pVQZ basis sets, plus CCSDTQ/cc-pVTZ $-$ CCSDT(Q)/cc-pVTZ difference\\
\end{flushleft}
\end{table}
\clearpage

\squeezetable
\begin{table}
\caption{Contribution of connected quintuple ($\hat{T}_5$) and sextuple ($\hat{T}_6$) excitations  to the total atomization energies (in kcal/mol)\label{tab:T5conv}}
\begin{tabular}{l|cc|ccc|cc|cc|cc}
\hline\hline
 & \multicolumn{2}{l}{~~~CCSDTQ5 $-$~~~} &  \multicolumn{3}{l}{~~~CCSDTQ(5) $-$~~~} & \multicolumn{2}{l}{~~~CCSDTQ5 $-$~~~} & \multicolumn{2}{l}{CCSDTQ56 $-$} & \multicolumn{2}{l}{~~~CCSDTQ5(6) $-$~~~} \\
 & \multicolumn{2}{r}{~~~~CCSDTQ} &  \multicolumn{3}{r}{~~~~CCSDTQ} & \multicolumn{2}{r}{~~~~CCSDTQ(5)} & \multicolumn{2}{r}{CCSDTQ5} & \multicolumn{2}{r}{CCSDTQ5} \\
Basis set & ~~DZ~~ & PVDZ & DZ & PVDZ & PVDZNOD & DZ & PVDZ & DZ & PVDZ & DZ & PVDZ \\
Reference & UHF & UHF & UHF & UHF & UHF & UHF & UHF & UHF & UHF & UHF & UHF \\
\hline
H$_2$O     & 0.01  & 0.01  & 0.01  & 0.01  & 0.01  & 0.00  & 0.00  &       &       & 0.00  & 0.00  \\   
B$_2$      & 0.08  & 0.08  & 0.07  & 0.05  & 0.08  & 0.02  & 0.03  &       &       & 0.00  & 0.00  \\    
C$_2$H$_2$ & 0.07  & 0.08  & 0.06  & 0.08  & 0.07  & 0.00  & 0.00  &       &       & 0.00  &       \\      
CH$_3$     & 0.00  & 0.00  & 0.00  & 0.00  & 0.00  & 0.00  & 0.00  &       &       & 0.00  & 0.00  \\   
CH$_4$     & 0.00  & 0.01  & 0.00  & 0.01  & 0.00  & 0.00  & 0.00  &       &       & 0.00  & 0.00  \\   
CH         & 0.00  & 0.00  & 0.00  & 0.00  & 0.00  & 0.00  & 0.00  &       &       & 0.00  & 0.00  \\    
CO$_2$     & 0.05  &       & 0.00  & -0.03 & 0.05  & 0.05  &       &       &       & 0.00  &       \\              
CO         & 0.05  & 0.03  & 0.06  & 0.02  & 0.05  & -0.01 & 0.01  & 0.00  &       & 0.00  & 0.00  \\
F$_2$      & 0.03  & 0.04  & 0.04  & 0.04  & 0.03  & -0.01 & 0.00  & 0.01  &       & 0.01  &       \\   
HF         & 0.01  & 0.00  & 0.01  & 0.00  & 0.01  & 0.00  & 0.00  & 0.00  &       & 0.00  & 0.00  \\
N$_2$      & 0.11  & 0.11  & 0.12  & 0.12  & 0.11  & -0.01 & -0.01 & 0.02  & 0.00  & 0.02  & 0.01  \\
NH$_3$     & 0.01  & 0.01  & 0.01  & 0.01  & 0.01  & 0.00  & 0.00  &       &       & 0.00  & 0.00  \\   
N$_2$O     & 0.19  &       & 0.04  & 0.06  & 0.19  & 0.14  &       &       &       & 0.02  &       \\               
NO         & 0.09  & 0.08  & 0.11  & 0.10  & 0.09  & -0.02 & -0.01 &       &       & 0.01  & 0.01  \\  
O$_2$      & 0.10  & 0.10  & 0.11  & 0.11  & 0.10  & 0.00  & -0.01 &       &       & 0.01  & 0.01  \\   
O$_3$      & 0.41  &       & 0.51  & 0.47  & 0.41  & -0.09 &       &       &       &       &       \\                               
C$_2$      & 0.27  & 0.32  & 0.47  & 0.47  & 0.27  & -0.20 & -0.15 & 0.07  & 0.06  & 0.07  & 0.06  \\
BN         & 0.18  & 0.16  & 0.11  & -0.11 & 0.18  & 0.07  & 0.27  & 0.04  & 0.02  & 0.07  & 0.01  \\
MgO        & -0.08 & -0.04 & -0.97 & -1.05 & -0.08 & 0.88  & 1.00  &       &       & -0.28 & -0.30 \\
BeO        & -0.13 & -0.11 & -0.72 & -0.72 & -0.13 & 0.59  & 0.62  & -0.03 &       & -0.15 & -0.16 \\
CN         & 0.13  & 0.12  & 0.16  & 0.14  & 0.13  & -0.03 & -0.03 &       &       & 0.01  &       \\      
NO$_2$     & 0.19  &       & 0.20  & 0.11  & 0.19  & -0.01 &       &       &       &       &       \\                              
Cl$_2$     & 0.00  & 0.02  & 0.00  & 0.02  & 0.00  & 0.00  & 0.00  &       &       & 0.00  & 0.00  \\   
ClF        & 0.02  & 0.02  & 0.02  & 0.02  & 0.02  & 0.00  & 0.00  &       &       & 0.00  & 0.00  \\   
CS         & 0.06  & 0.05  & 0.05  & 0.01  & 0.06  & 0.00  & 0.04  & 0.00  &       & 0.00  & 0.00  \\
H$_2$S     & 0.00  & 0.00  & 0.00  & 0.00  & 0.00  & 0.00  & 0.00  &       &       & 0.00  & 0.00  \\   
HCl        & 0.00  & 0.00  & 0.00  & 0.00  & 0.00  & 0.00  & 0.00  &       &       & 0.00  & 0.00  \\   
HOCl       & 0.02  &       & 0.02  & 0.03  & 0.02  & 0.00  &       &       &       & 0.00  &       \\              
PH$_3$     & 0.00  & 0.00  & 0.00  & 0.00  & 0.00  & 0.00  & 0.00  &       &       & 0.00  & 0.00  \\   
SO         & 0.05  & 0.06  & 0.05  & 0.06  & 0.05  & 0.00  & 0.00  &       &       & 0.01  & 0.01  \\    
SO$_2$     &       &       & 0.14  &       & 0.14  & 0.00  &       &       &       &       &       \\                               
OCS        &       &       & -0.01 & -0.04 & 0.05  & 0.00  &       &       &       &       &       \\                        
ClCN       & 0.11  &       & 0.11  & 0.12  & 0.11  & 0.00  &       &       &       &       &       \\                          
C$_2$H$_4$ & 0.03  &       & 0.02  & 0.03  & 0.03  & 0.00  &       &       &       & 0.00  &       \\              
H$_2$CO    & 0.03  &       & 0.04  & 0.03  & 0.03  & -0.01 &       &       &       & 0.00  &       \\             
HNO        & 0.07  & 0.08  & 0.08  & 0.09  & 0.07  & -0.01 & -0.01 &       &       & 0.01  &       \\     
\hline\hline
\end{tabular}
\end{table}
\clearpage

\squeezetable
\begin{table}
\caption{Basis set convergence of CCSDT$-$CCSD(T); post-CCSD(T) core correlation contributions calculated with the cc-pCVTZ basis set ($\Delta$TAE, in kcal/mol)\label{tab:T3conv}\label{tab:coreT3T4} and basis set convergence of CCSD(T) MOLPRO$-$ACESII differences in the TAE (kcal/mol) \label{tab:ACESvsMOLPRO}}
\begin{tabular}{l|ccc|ccc|ccccc}
\hline\hline
& \multicolumn{3}{c}{CCSDT$-$CCSD(T) (a)} & \multicolumn{3}{c}{Post-CCSD(T) core corr.} & \multicolumn{5}{c}{CCSD(T) MOLPRO$-$ACESII diff.}\\
&  $\{$PVDZ, & $\{$PVTZ, & diff. &  ACES & MOLPRO &	& PVDZ 	&	PVTZ 	&	PVQZ 	&	$\{$PVDZ,	&\ $\{$PVTZ,\\
& PVTZ$\}$ & PVQZ$\}$ & & $\Delta$$\Delta$[$\hat{T}_3$-(T)] & $\Delta$$\Delta$[$\hat{T}_3$-(T)] & $\Delta$$\Delta$(Q) & & & & PVTZ$\}$ & PVQZ$\}$ \\
\hline
H$_2$O     & -0.18 & -0.20 & -0.03 & -0.02 &  0.01 &  0.00 &  0.02 &  0.02 &  0.02 &  0.02 &  0.02 \\
B$_2$      &  0.18 &  0.07 & -0.11 &  0.04 &  0.04 &  0.07 &  0.00 &  0.00 &  0.00 &  0.00 &  0.00 \\
C$_2$H$_2$ & -0.68 & -0.68 &  0.00 & -0.01 &  0.03 &  0.01 &  0.03 &  0.04 &  0.04 &  0.04 &  0.04 \\
CH$_3$     & -0.01 & -0.03 & -0.01 & -0.01 &  0.01 &  0.00 &  0.01 &  0.02 &  0.02 &  0.02 &  0.02 \\
CH$_4$     & -0.06 & -0.07 & -0.01 & -0.01 &  0.01 &  0.00 &  0.01 &  0.02 &  0.02 &  0.02 &  0.02 \\
CH         &  0.12 &  0.12 &  0.00 & -0.01 &  0.01 &  0.00 &  0.01 &  0.01 &  0.02 &  0.02 &  0.02 \\
CO$_2$     & -0.96 & -0.99 & -0.03 & -0.03 &  0.04 &       &  0.05 &  0.06 &  0.07 &  0.07 &  0.07 \\
CO         & -0.51 & -0.51 &  0.00 & -0.02 &  0.02 &  0.02 &  0.03 &  0.04 &  0.04 &  0.04 &  0.05 \\
F$_2$      & -0.34 & -0.29 &  0.05 &  0.02 &  0.03 &  0.01 &  0.01 &  0.01 &  0.01 &  0.01 &  0.01 \\
HF         & -0.13 & -0.15 & -0.02 & -0.01 &  0.00 &  0.00 &  0.00 &  0.01 &  0.01 &  0.01 &  0.01 \\
N$_2$      & -0.69 & -0.66 &  0.02 & -0.05 &  0.04 &  0.01 &  0.07 &  0.09 &  0.10 &  0.10 &  0.10 \\
NH$_3$     & -0.09 &       &       & -0.03 &  0.01 &  0.00 &  0.03 &  0.05 &  0.05 &  0.05 &  0.05 \\
N$_2$O     & -1.38 &       &       & -0.05 &  0.07 &       &  0.08 &  0.11 &  0.12 &  0.12 &  0.13 \\
NO         & -0.49 & -0.47 &  0.02 & -0.03 &  0.03 &  0.02 &  0.05 &  0.06 &  0.07 &  0.07 &  0.07 \\
O$_2$      & -0.71 & -0.64 &  0.07 &  0.00 &  0.03 &  0.01 &  0.02 &  0.03 &  0.03 &  0.03 &  0.03 \\
O$_3$      & -1.27 &       &       &  0.06 &  0.13 &       &  0.05 &  0.06 &  0.07 &  0.07 &  0.07 \\
C$_2$      & -2.15 & -2.19 & -0.05 &  0.15 &  0.19 &  0.08 &  0.03 &  0.04 &  0.04 &  0.04 &  0.04 \\
BN         & -2.59 & -2.58 &  0.01 &  0.07 &  0.12 &  0.12 &  0.04 &  0.05 &  0.05 &  0.05 &  0.06 \\
CN         & -0.28 & -0.27 &  0.01 & -0.03 &  0.03 &  0.03 &  0.05 &  0.07 &  0.07 &  0.07 &  0.07 \\
NO$_2$     & -0.98 &       &       & -0.02 &  0.07 &       &  0.07 &  0.09 &  0.09 &  0.10 &  0.10 \\
Cl$_2$     & -0.35 & -0.37 & -0.02 & -0.02 & -0.01 &       &  0.01 &  0.02 &  0.02 &  0.02 &  0.02 \\
ClF        & -0.28 & -0.27 &  0.01 &  0.01 &  0.03 &       &  0.01 &  0.01 &  0.02 &  0.01 &  0.02 \\
CS         & -0.58 & -0.57 &  0.01 & -0.05 & -0.02 &       &  0.02 &  0.03 &  0.03 &  0.03 &  0.04 \\
H$_2$S     & -0.08 & -0.07 &  0.01 &  0.00 &  0.01 &       &  0.01 &  0.01 &  0.01 &  0.01 &  0.01 \\
HCl        & -0.10 & -0.11 & -0.01 & -0.01 &  0.00 &       &  0.01 &  0.01 &  0.01 &  0.01 &  0.01 \\
HOCl       & -0.44 &       &       &       &       &       &  0.02 &  0.03 &  0.03 &  0.03 &  0.03 \\
PH$_3$     & -0.02 &  0.07 &  0.08 &  0.02 &  0.03 &       &  0.01 &  0.01 &  0.01 &  0.01 &  0.01 \\
SO         & -0.76 & -0.74 &  0.02 &  0.02 &  0.04 &       &  0.02 &  0.02 &  0.03 &  0.03 &  0.03 \\
SO$_2$     & -1.22 &       &       &       &       &       &  0.04 &  0.05 &  0.06 &  0.06 &  0.06 \\
OCS        & -1.04 &       &       &       &       &       &  0.04 &  0.05 &  0.06 &  0.06 &  0.06 \\
ClCN       & -1.07 &       &       &       &       &       &  0.05 &  0.07 &  0.08 &  0.08 &  0.08 \\
C$_2$H$_4$ & -0.42 & -0.43 & -0.01 & -0.01 &  0.03 &  0.00 &  0.03 &  0.04 &  0.04 &  0.04 &  0.04 \\
H$_2$CO    & -0.48 & -0.49 & -0.01 & -0.02 &  0.02 &  0.01 &  0.03 &  0.04 &  0.04 &  0.04 &  0.05 \\
HNO        & -0.52 & -0.48 &  0.04 & -0.03 &  0.04 &  0.01 &  0.05 &  0.07 &  0.07 &  0.07 &  0.08 \\
\hline\hline
\end{tabular}
\begin{flushleft}
(a) All values were calculated using ACESII 
\end{flushleft}
\end{table}
\clearpage

\squeezetable
\begin{table}

\caption{Basis set convergence of diagonal Born-Oppenheimer corrections at the Hartree-Fock level and of scalar relativistic corrections (TAE, in kcal/mol)  \label{tab:DBOC}}
\begin{tabular}{l|cc|ccc}
\hline\hline
    & \multicolumn{2}{c}{DBOC} & \multicolumn{3}{c}{Scalar relativistic} \\
	&	 &  & Oren & PNNL & PNNL	\\
	&	AVDZ & AVTZ & AVQZ & AV(T+d)Z & AV(Q+d)Z	\\
\hline
H$_2$O & 0.12 & 0.13  &  -0.26  & -0.27 & -0.27  \\
B$_2$ & 0.01 & 0.01   &  -0.06  & -0.06 & -0.06  \\
C$_2$H$_2$ & 0.12 & 0.12 & -0.28        & -0.28 & -0.28  \\
CH$_3$ & 0.05 & 0.05   & -0.17  & -0.17 & -0.17  \\
CH$_4$ & 0.11 & 0.10   & -0.19  & -0.19 & -0.19  \\
CH & -0.08 & -0.08  &    -0.04  & -0.04 & -0.04  \\
CO$_2$ & 0.05 & 0.05  &  -0.47  & -0.48 & -0.48  \\
CO & 0.01 & 0.02  &      -0.16  & -0.16 & -0.16  \\
F$_2$ & 0.00 & 0.00  &   -0.02  & -0.03 & -0.03  \\
HF & 0.08 & 0.08  &      -0.19  & -0.20 & -0.20  \\
N$_2$ & 0.02 & 0.02  &   -0.13  & -0.14 & -0.14  \\
NH$_3$ & 0.14 & 0.14  &  -0.25  & -0.26 & -0.25  \\
N$_2$O & 0.04 & 0.04  &  -0.45  & -0.46 & -0.46  \\
NO & 0.01 & 0.01  &      -0.18  & -0.19 & -0.19  \\
O$_2$ & 0.01 & 0.01  &   -0.18  & -0.18 & -0.18  \\
O$_3$ & -0.03 & -0.03  & -0.24  & -0.25 & -0.25  \\
C$_2$ & 0.03 & 0.03  &   -0.17  & -0.17 & -0.17  \\
BN & 0.01 & 0.01  &      -0.17  & -0.17 & -0.17  \\
CN & 0.02 & 0.02  &      -0.15  & -0.16 & -0.16  \\
NO$_2$ & 0.00 & 0.00  &  -0.42  & -0.43 & -0.43  \\
Cl$_2$ & 0.00 & 0.00  &  -0.21  & -0.19 & -0.20  \\
ClF & 0.00 & 0.00  &     -0.18  & -0.17 & -0.18  \\
CS & 0.01 & 0.01  &      -0.16  & -0.16 & -0.16  \\
H$_2$S & 0.06 & 0.05  &  -0.39  & -0.39 & -0.40  \\
HCl & 0.05 & 0.04  &     -0.25  & -0.25 & -0.25  \\
HOCl & 0.07 & 0.07  &    -0.32  & -0.32 & -0.33  \\
PH$_3$ & 0.16 & 0.15  &  -0.45  & -0.45 & -0.46  \\
SO & 0.01 & 0.01  &      -0.34  & -0.33 & -0.34  \\
SO$_2$ & 0.02 & 0.02  &  -0.81  & -0.82 & -0.83  \\
OCS & 0.03 & 0.03  &     -0.53  & -0.54 & -0.54  \\
ClCN & 0.04 & 0.04  &    -0.44  & -0.44 & -0.45  \\
C$_2$H$_4$ & 0.12 & 0.12 & -0.33        & -0.33 & -0.33  \\
H$_2$CO & 0.03 & 0.03  & -0.33  & -0.34 & -0.34  \\
HNO & -0.05 & -0.05  &   -0.27  & -0.27 & -0.27  \\
\hline\hline
\end{tabular}
\end{table}
\clearpage

\squeezetable
\begin{table}
\caption{Component breakdown of the final W4 total atomization energies at the bottom of the well (in kcal/mol)\label{tab:W4components}}
\begin{tabular}{l|cccccccccccc}
\hline\hline
 & SCF & valence & valence & $\hat{T}_3-$(T) & $\hat{T}_4$ & $\hat{T}_5$ & inner & relativ. & spin-orbit & DBOC & (a) & TAE$_e$ \\
 & & CCSD & (T) & & & & shell &  &  &  &  &  \\
\hline
H$_2$O     & 160.02 & 69.08  & 3.53  & -0.20 & 0.18 & 0.01 & 0.38  & -0.27 & -0.22 & 0.13  & 0.02 & 232.65  \\
C$_2$H$_2$ & 299.87 & 94.73  & 8.35  & -0.72 & 0.70 & 0.07 & 2.49  & -0.28 & -0.17 & 0.12  & 0.04 & 405.18  \\
CH$_3$     & 243.40 & 61.47  & 1.90  & -0.03 & 0.05 & 0.00 & 1.09  & -0.17 & -0.08 & 0.05  & 0.02 & 307.69  \\
CH$_4$     & 331.55 & 84.72  & 2.89  & -0.08 & 0.08 & 0.00 & 1.27  & -0.19 & -0.08 & 0.10  & 0.02 & 420.26  \\
CH         & 57.22  & 25.83  & 0.89  & 0.10  & 0.03 & 0.00 & 0.14  & -0.04 & -0.04 & -0.08 & 0.02 & 84.06   \\
CO$_2$     & 258.08 & 116.34 & 13.86 & -1.03 & 1.04 & 0.05 & 1.77  & -0.48 & -0.53 & 0.05  & 0.07 & 389.18  \\
CO         & 181.58 & 69.08  & 8.02  & -0.55 & 0.61 & 0.05 & 0.96  & -0.16 & -0.31 & 0.02  & 0.04 & 259.30  \\
F$_2$      & -31.08 & 61.91  & 7.63  & -0.35 & 0.89 & 0.03 & -0.10 & -0.03 & -0.77 & 0.00  & 0.01 & 38.14   \\
HF         & 100.05 & 39.31  & 2.15  & -0.13 & 0.10 & 0.01 & 0.18  & -0.20 & -0.39 & 0.08  & 0.01 & 141.18  \\
N$_2$      & 119.69 & 98.11  & 9.49  & -0.79 & 1.03 & 0.11 & 0.79  & -0.14 & 0.00  & 0.02  & 0.10 & 228.37  \\
NH$_3$     & 203.28 & 90.17  & 3.89  & -0.14 & 0.16 & 0.01 & 0.65  & -0.25 & 0.00  & 0.14  & 0.05 & 297.93  \\
N$_2$O     & 95.13  & 155.03 & 18.81 & -1.51 & 1.98 & 0.19 & 1.15  & -0.46 & -0.22 & 0.04  & 0.12 & 270.20  \\
NO         & 54.92  & 87.51  & 9.48  & -0.56 & 0.84 & 0.09 & 0.41  & -0.19 & -0.05 & 0.01  & 0.07 & 152.52  \\
O$_2$      & 26.78  & 83.95  & 9.28  & -0.74 & 1.07 & 0.10 & 0.23  & -0.18 & -0.45 & 0.01  & 0.03 & 120.15  \\
O$_3$      & -45.09 & 163.94 & 25.62 & -1.34 & 3.81 & 0.41 & -0.05 & -0.25 & -0.67 & -0.03 & 0.07 & 146.39  \\
NO$_2$     & 59.55  & 147.16 & 19.36 & -1.08 & 1.91 & 0.19 & 0.67  & -0.43 & -0.45 & 0.00  & 0.10 & 226.99  \\
Cl$_2$     & 26.82  & 28.08  & 4.59  & -0.37 & 0.44 & 0.00 & 0.14  & -0.20 & -1.68 & 0.00  & 0.02 & 57.83   \\
ClF        & 15.41  & 41.85  & 5.25  & -0.30 & 0.46 & 0.02 & 0.04  & -0.18 & -1.23 & 0.00  & 0.01 & 61.35   \\
CS         & 104.16 & 57.06  & 9.68  & -0.62 & 0.98 & 0.06 & 0.84  & -0.16 & -0.64 & 0.01  & 0.03 & 171.38  \\
H$_2$S     & 133.63 & 47.66  & 2.24  & -0.10 & 0.15 & 0.00 & 0.33  & -0.40 & -0.56 & 0.05  & 0.01 & 183.00  \\
HCl        & 80.85  & 24.99  & 1.48  & -0.11 & 0.10 & 0.00 & 0.19  & -0.25 & -0.84 & 0.04  & 0.01 & 106.47  \\
HOCl       & 86.70  & 72.35  & 6.77  & -0.47 & 0.58 & 0.02 & 0.26  & -0.33 & -1.06 & 0.07  & 0.03 & 164.90  \\
PH$_3$     & 173.22 & 66.54  & 2.01  & -0.03 & 0.09 & 0.00 & 0.34  & -0.46 & 0.00  & 0.15  & 0.01 & 241.88  \\
SO         & 53.14  & 64.15  & 8.46  & -0.79 & 0.84 & 0.05 & 0.49  & -0.34 & -0.78 & 0.01  & 0.03 & 125.30  \\
SO$_2$     & 121.91 & 121.45 & 15.83 & -1.28 & 1.56 & 0.14 & 0.97  & -0.83 & -1.01 & 0.02  & 0.06 & 258.80  \\
OCS        & 218.25 & 101.32 & 14.47 & -1.09 & 1.33 & 0.05 & 1.41  & -0.54 & -0.87 & 0.03  & 0.06 & 334.37  \\
ClCN       & 169.49 & 101.51 & 12.44 & -1.15 & 1.24 & 0.11 & 1.78  & -0.45 & -0.93 & 0.04  & 0.08 & 284.12  \\
C$_2$H$_4$ & 434.97 & 119.32 & 7.40  & -0.46 & 0.43 & 0.03 & 2.38  & -0.33 & -0.17 & 0.12  & 0.04 & 563.71  \\
H$_2$CO    & 264.83 & 100.53 & 7.92  & -0.53 & 0.54 & 0.03 & 1.31  & -0.34 & -0.31 & 0.03  & 0.04 & 374.04  \\
HNO        & 85.45  & 109.46 & 10.08 & -0.60 & 0.99 & 0.07 & 0.40  & -0.27 & -0.22 & -0.05 & 0.07 & 205.34  \\
\hline\hline
\end{tabular}
\begin{flushleft}
(a) difference between the ACES II and MOLPRO definitions of the valence ROCCSD(T)
\end{flushleft}
\end{table}
\clearpage

\squeezetable
\begin{table}
\caption{Comparison between W4 total atomization energies at 0 K, Active Thermochemical Tables benchmarks, and earlier reference data (kcal/mol)\label{tab:W4results}}
\begin{tabular}{l|cccc|c|cc|cc|cc}
\hline\hline
         & ZPVE(a) &  W2.2 &  W3.2  & W4lite &   W4    & W4.2   &  W4.3  &ATcT(b)&uncert. &Earlier ref. (c) & uncert.  \\ \hline
H$_2$O      & 13.29 & 219.39 & 219.47 & 219.43 &  219.36 & 219.35 & 219.35 & 219.36 & 0.01 & 219.35 & 0.12  \\
C$_2$H$_2$  & 16.46 & 388.61 & 388.54 & 388.57 &  388.72 & 388.72 &        & 388.62 & 0.07 & 388.90 & 0.24  \\
CH$_3$      & 18.55 & 289.12 & 289.15 & 289.14 &  289.14 & 289.14 &        & 289.11 & 0.03 & 289.00 & 0.10  \\
CH$_4$      & 27.74 & 392.52 & 392.52 & 392.52 &  392.52 & 392.52 &        & 392.50 & 0.03 & 392.51 & 0.14  \\
CH          &  4.04 &  79.89 &  80.01 &  80.01 &   80.02 &  80.02 &  80.02 &  79.98 & 0.05 & 79.90 & 0.23  \\
CO$_2$      &  7.24 & 381.70 & 381.94 & 382.06 &  381.94 & 381.94 &        & 382.01 & 0.03 & 381.91 & 0.06  \\
CO          &  3.11 & 256.01 & 256.12 & 256.17 &  256.19 & 256.18 & 256.21 & 256.25 & 0.03 & 256.16 & 0.12  \\
F$_2$       &  1.30 &  36.15 &  36.78 &  36.85 &   36.84 &  36.87 &  36.97 &  36.91 & 0.07 & 36.94 & 0.10  \\
HF          &  5.85 & 135.32 & 135.39 & 135.40 &  135.33 & 135.32 & 135.30 & 135.27 & 0.00 & 135.33 & 0.17  \\
N$_2$       &  3.36 & 224.62 & 224.92 & 224.90 &  225.01 & 225.00 & 225.07 & 224.94 & 0.01 & 225.06 & 0.04  \\
NH$_3$      & 21.33 & 276.62 & 276.68 & 276.62 &  276.60 & 276.59 & 276.61 & 276.59 & 0.01 & 276.73 & 0.13  \\
N$_2$O      &  6.81 & 262.44 & 263.24 & 263.42 &  263.39 & 263.40 &        & 263.38 & 0.03 & 263.79 &   \\
NO          &  2.71 & 149.30 & 149.66 & 149.74 &  149.81 & 149.81 & 149.86 & 149.82 & 0.02 & 149.82 & 0.03  \\
O$_2$       &  2.25 & 117.25 & 117.69 & 117.77 &  117.90 & 117.91 & 118.00 & 117.99 & 0.00 & 117.97 & 0.04  \\
O$_3$       &  4.15 & 139.06 & 142.06 & 142.15 &  142.24 & 142.34 &        & 142.48 & 0.01 & 142.51 &   \\
NO$_2$      &  5.40 & 220.24 & 221.31 & 221.49 &  221.59 & 221.61 &        & 221.67 & 0.02 & 221.70 &   \\
Cl$_2$      &  0.80 &  57.12 &  57.03 &  56.85 &   57.03 &  57.01 &  57.08 &  57.18 & 0.00 & 57.18 & 0.00  \\
ClF         &  1.12 &  60.00 &  60.17 &  60.19 &   60.23 &  60.24 &  60.29 &  &  & 60.36 & 0.01  \\
CS          &  1.83 & 169.24 & 169.24 & 169.10 &  169.55 & 169.51 & 169.59 &  &  & 169.41 & 0.23  \\
H$_2$S      &  9.40 & 173.59 & 173.58 & 173.54 &  173.60 & 173.60 & 173.64 & 173.55 & 0.07 & 173.15 & 0.12  \\
HCl         &  4.24 & 102.32 & 102.28 & 102.20 &  102.23 & 102.22 & 102.23 & 102.21 & 0.00 & 102.24 & 0.02  \\
HOCl        &  8.18 & 156.61 & 156.71 & 156.67 &  156.72 &        &        & 156.64 & 0.43 & 156.61 & 0.12  \\
PH$_3$      & 14.44 & 227.36 & 227.39 & 227.40 &  227.44 & 227.47 & 227.57 &  &  & 227.13 & 0.41  \\
SO          &  1.64 & 123.47 & 123.53 & 123.52 &  123.66 & 123.69 & 123.76 & 123.72 & 0.02 & 123.58 & 0.04  \\
SO$_2$      &  4.38 & 253.75 & 254.37 & 254.53 &  254.42 &        &        & 254.46 & 0.02 & 253.92 & 0.08  \\
OCS         &  5.72 & 328.40 & 328.52 & 328.45 &  328.65 &        &        &  &  & 328.53 & 0.48  \\
ClCN        &  5.33 & 278.73 & 278.77 & 278.58 &  278.79 &        &        & 279.42 & 0.26 & 279.20 & 0.48  \\
C$_2$H$_4$  & 31.60 & 532.06 & 531.99 & 532.02 &  532.11 & 532.11 &        & 532.00 & 0.06 & 531.91 & 0.17  \\
H$_2$CO     & 16.53 & 357.38 & 357.48 & 357.53 &  357.51 & 357.51 &        & 357.43 & 0.06 & 357.25 & 0.12  \\
HNO         &  8.56 & 196.27 & 196.71 & 196.71 &  196.78 & 196.78 &        & 196.86 & 0.03 & 196.85 & 0.06  \\
\hline\hline
\end{tabular}
\begin{flushleft}
(a) Zero-point vibrational energies taken from the compilation in Ref.\cite{w1}, except for the following:\\
{\bf CH$_3$}: D. W. Schwenke, {\em Spectrochimica Acta A} {\bf 55}, 731 (1999), adjusted for expt.--calc. differences in the fundamental frequencies;
{\bf CH$_4$}: D. W. Schwenke, {\em Spectrochimica Acta A} {\bf 58}, 849 (2002), same adjustment made;
{\bf H$_2$O}: CCSD(T)/cc-pVQZ quartic force field, fundamentals obtained by second-order rovibrational perturbation theory and 
adjusted for expt.-calc. differences in fundamentals;
{\bf HOCl}: K. A. Peterson, S. Skokov, and J. M. Bowman, {\em J. Chem. Phys.} {\bf 111}, 7446 (1999);
{\bf N$_2$O}: as for H$_2$O;
{\bf O$_3$}: Obtained from experimentally derived harmonic frequencies and anharmonicity constants in 
A. Barbe, A. Chichery, T. Cours, V. G. Tyuterev, and J. J. Plateaux,
{\em J. Mol. Struct.} {\bf 616}, 55 (2002);
{\bf NO$_2$}:  from R. Georges, A. Delon, and R. Jost,
{\em J. Chem. Phys.} {\bf 103}, 1732 (1995); number not explicitly given there, 
quoted (presumably derived from their potential) by
A. J. C. Varandas, {\em J. Chem. Phys.} {\bf 119}, 2596 (2003).

(b) Ref.\cite{Branko5}, except for H$_2$S, SO, and SO$_2$, which are from Ref.\cite{Branko5b};  the adjunct uncertainties correspond to 95\% confidence
intervals, as customary in experimental thermochemistry, which were
obtained by utilizing the full covariance matrix computed by ATcT; see
see also Refs.\cite{Branko1,Branko2,Branko3,Branko4}.

(c) Ref.\cite{w3} and references therein.
\end{flushleft}
\end{table}
\clearpage

\squeezetable
\begin{table}
\caption{Components of W1w, W2w, W3, W2.2, W3.2, W4lite, W4, W4.2 and W4.3 \label{tab:WNdefs}}
\begin{tabular}{l | c | c | c | c | c | c | c | c | c |}
\hline\hline
  Component         & W1w & W2w & W3 & W2.2 &  W3.2 &  W4lite  & W4 &   W4.2    & W4.3   \\ \hline
Reference geometry	& B3LYP/ & \multicolumn{8}{c}{CCSD(T)/cc-pV(Q+d)Z}\\
                    & cc-pV(T+d)Z &&&&&&&&\\
\hline
SCF					& AV\{T,Q\}+dZ &\multicolumn{3}{c}{~~~~~~~~~AV\{Q,5\}+dZ}& &   \multicolumn{4}{c}{AV\{5,6\}+dZ} \\
\hline
SCF extrap.         &    (a) & (a) & (a) & (b) & (b) & (b) & (b) & (b) & (b) \\
\hline
Valence CCSD			& AV\{T,Q\}+dZ &\multicolumn{3}{c}{~~~~~~~~~AV\{Q,5\}+dZ}& & \multicolumn{4}{c}{AV\{5,6\}+dZ} \\
\hline
Val. CCSD extrap.   &        (c)     &  (d)   &  (d) &  (e)  & (e)   & (e) & (e) & (e)    & (e)    \\
\hline
Valence (T)			& AV\{D,T\}+dZ &\multicolumn{3}{c}{~~~~~~~~~AV\{T,Q\}+dZ}&  &\multicolumn{4}{c}{AV\{Q,5\}+dZ} \\
\hline
Val. (T) extrap.    &        (c)      &  (d)   & (d)   &  (d)   &  (d)   &  (d)   &  (d)   &  (d)   &  (d)    \\
\hline
Valence 	$T_3$-(T)		& ---  & --- &	V\{D,T\}Z (f) & --- & \multicolumn{3}{c}{~~~~~~V\{D,T\}Z (f)} && V\{T,Q\}Z (f)\\
\hline
Valence (Q)			& --- & --- &	1.25 PVDZ  & --- & PVDZ & PVDZ & \multicolumn{2}{c}{1.10 PVTZ} & PVQZ \\
\hline
Valence $T_4$-(Q)	& --- & --- &	1.25 PVDZ  & --- & ---  & --- & \multicolumn{2}{c}{1.10 PVDZ} & PVTZ \\
\hline
Valence $T_5$	    & --- & --- & ---  & ---  & ---  & --- & DZ (f) & DZ (f) & PVDZ \\
\hline
CCSD(T) inner shell	& \multicolumn{2}{c}{~~~~~~~~~MTsmall} && \multicolumn{6}{c}{ACV\{T,Q\}Z} \\
\hline
$T_3$-(T) inner shell	& --- & --- & --- & --- & --- & --- & --- & CVTZ & CVTZ\\
\hline
Scalar relativistics& \multicolumn{2}{c}{~~~~~~~~~MTsmall} && \multicolumn{6}{c}{AVQ+dZ} \\
\hline
ZPVE					& 0.985 B3LYP/ & (h) & (h) & (i) & (i) & (i) & (i) & (i) & (i)\\
     				& cc-pV(T+d)Z  &&&&&&&&\\
\hline
DBOC					& --- & --- & --- &\multicolumn{6}{c}{HF/AVTZ}\\
\hline\hline
\end{tabular}
\begin{flushleft}
The notation \{D,T\} refers to extrapolation from, in this case, cc-pVDZ and cc-pVTZ basis sets.

(a) $A+B/L^5$\\
(b) Martin-Karton formula\cite{MKtca}.\\
(c) $A+B/L^{3.22}$ on valence correlation energies.\\
(d) $A+B/L^3$ on valence correlation energies.\\
(e) Separate $L^{-3}$ and $L^{-5}$ extrapolations on singlet and triplet coupled pairs, respectively.\\
(f) CCSDT calculated with ACES II, CCSD(T) calculated with MOLPRO.\\
(g) For second row systems a cc-pVDZ basis without the d functions was used.\\
(h) From CCSD(T)/cc-pV(Q+d)Z quartic force field.\\
(i) Best available, usually obtained from expt. fundamentals and high-level ab initio anharmonic force field (whence anharmonic corrections). See Table \ref{tab:W4results} for further details.

\end{flushleft}
\end{table}
\clearpage

\begin{table}

\caption{Valence-only and all-electrons-correlated total atomization energies (kcal/mol) at the CCSDT level with different reference orbital choices, as well as using different definitions of CCSD(T)\label{tab:A1}. The cc-pCVTZ basis set
was used throughout.\label{tab:TABLEV}}
\begin{tabular}{l|cccccc|cccccc}
\hline\hline
 & CCSD(T) & CCSD(T) & CCSD(T) & CCSDT & CCSDT & CCSDT & CCSD(T) & CCSD(T) & CCSD(T) & CCSDT & CCSDT & CCSDT  \\
Ref. & UHF & ROHF & ROHF & ROHF & UHF & ROHF & UHF & ROHF & ROHF & ROHF & UHF & ROHF  \\    
Def. & ~ & Aces & MOLPRO & ~ & ~ & ~ & ~ & Aces & MOLPRO & ~ & ~ & ~ \\    
Orbs. &  &  &  & semican. & std. & std. &  &  &  & semican. & std. & std.  \\
\hline
& \multicolumn{6}{c}{All electrons correlated}  & \multicolumn{6}{c}{Only valence electrons correlated}  \\
\hline
B$_2$ & 63.38 & 63.44 & 63.44 & 63.76 & 63.75 & 63.76 & 62.85 & 62.93 & 62.93 & 63.21 & 63.19 & 63.21  \\
BN & 98.56 & 98.57 & 98.57 & 96.18 & 96.18 & 96.18 & 97.72 & 97.82 & 97.87 & 95.31 & 95.22 & 95.37  \\
C$_2$ & 140.37 & 140.37 & 140.37 & 138.62 & 138.61 & 138.62 & 139.51 & 139.60 & 139.64 & 137.66 & 137.56 & 137.69  \\
N$_2$ & 217.93 & 217.94 & 217.94 & 217.37 & 217.36 & 217.37 & 217.00 & 217.17 & 217.27 & 216.56 & 216.39 & 216.65  \\
CO & 253.03 & 253.03 & 253.03 & 252.64 & 252.64 & 252.64 & 252.11 & 252.23 & 252.27 & 251.82 & 251.70 & 251.86  \\
CO$_2$ & 377.86 & 377.86 & 377.86 & 377.08 & 377.08 & 377.08 & 376.18 & 376.36 & 376.43 & 375.55 & 375.36 & 375.61  \\
N$_2$O & 255.66 & 255.67 & 255.67 & 254.49 & 254.49 & 254.49 & 254.23 & 254.47 & 254.59 & 253.23 & 252.99 & 253.34  \\
O$_2$ & 114.35 & 114.38 & 114.38 & 113.85 & 113.98 & 113.85 & 113.97 & 114.08 & 114.12 & 113.52 & 113.57 & 113.55  \\
O$_3$ & 133.25 & 133.25 & 133.25 & 132.36 & 132.36 & 132.36 & 132.86 & 133.06 & 133.13 & 132.05 & 131.84 & 132.12  \\
F$_2$ & 35.05 & 35.05 & 35.05 & 34.85 & 34.85 & 34.85 & 35.01 & 35.09 & 35.10 & 34.86 & 34.77 & 34.87  \\
HF & 137.43 & 137.43 & 137.43 & 137.33 & 137.33 & 137.33 & 137.21 & 137.25 & 137.25 & 137.15 & 137.11 & 137.16  \\
CN & 171.22 & 172.39 & 172.39 & 172.25 & 172.26 & 172.25 & 170.20 & 171.46 & 171.53 & 171.29 & 171.17 & 171.36  \\
NO & 144.19 & 144.35 & 144.35 & 143.99 & 144.02 & 143.99 & 143.59 & 143.88 & 143.95 & 143.48 & 143.39 & 143.55  \\
NO$_2$ & 213.37 & 213.50 & 213.50 & 212.77 & 212.83 & 212.77 & 212.36 & 212.70 & 212.79 & 211.90 & 211.75 & 211.99  \\
CH & 81.94 & 81.94 & 81.94 & 82.05 & 82.05 & 82.05 & 81.77 & 81.80 & 81.81 & 81.90 & 81.88 & 81.92  \\  
CH$_3$ & 303.51 & 303.52 & 303.52 & 303.51 & 303.51 & 303.51 & 302.63 & 302.66 & 302.68 & 302.65 & 302.63 & 302.67  \\
C$_2$H$_2$ & 396.08 & 396.08 & 396.08 & 395.53 & 395.53 & 395.53 & 394.09 & 394.18 & 394.22 & 393.61 & 393.52 & 393.65  \\
C$_2$H$_4$ & 554.36 & 554.36 & 554.36 & 554.04 & 554.04 & 554.04 & 552.36 & 552.45 & 552.49 & 552.12 & 552.02 & 552.15  \\
CH$_4$ & 415.31 & 415.31 & 415.31 & 415.27 & 415.27 & 415.27 & 414.23 & 414.28 & 414.30 & 414.23 & 414.18 & 414.25  \\
NH$_3$ & 289.79 & 289.79 & 289.79 & 289.72 & 289.72 & 289.72 & 289.07 & 289.16 & 289.21 & 289.08 & 289.00 & 289.13  \\
H$_2$O & 226.13 & 226.13 & 226.13 & 225.99 & 225.99 & 225.99 & 225.68 & 225.75 & 225.77 & 225.60 & 225.53 & 225.63  \\
H$_2$CO & 365.67 & 365.67 & 365.67 & 365.29 & 365.29 & 365.29 & 364.43 & 364.54 & 364.59 & 364.15 & 364.04 & 364.20  \\
HNO & 195.73 & 195.73 & 195.73 & 195.34 & 195.34 & 195.34 & 195.08 & 195.24 & 195.31 & 194.81 & 194.65 & 194.88  \\
Cl$_2$  &  53.93 &  53.91 & 53.91  &  53.62 &  53.61 & 53.62  &  53.60 &  53.67 & 53.69  &  54.19 &  53.29 & 53.41  \\
ClF     &  56.51 &  56.50 & 56.50  &  56.32 &  56.31 & 56.32  &  56.31 &  56.38 & 56.40  &  56.57 &  56.08 & 56.19  \\
CS      & 164.90 & 164.89 & 164.89 & 164.45 & 164.43 & 164.45 & 164.06 & 164.17 & 164.21 & 164.23 & 164.61 & 163.77  \\
H$_2$S  & 179.69 & 179.69 & 179.69 & 179.64 & 179.63 & 179.64 & 179.30 & 179.37 & 179.38 & 179.80 & 179.24 & 179.33  \\
HCl     & 105.14 & 105.13 & 105.13 & 105.04 & 105.03 & 105.04 & 104.91 & 104.94 & 104.95 & 105.26 & 104.81 & 104.87  \\
PH$_3$  & 236.74 & 236.74 & 236.74 & 236.81 & 236.81 & 236.81 & 236.35 & 236.45 & 236.46 & 236.97 & 236.38 & 236.49  \\
SO      & 118.40 & 118.63 & 118.63 & 118.07 &    N/A & 118.07 & 117.84 & 118.18 & 118.20 & 118.06 & 117.53 & 117.60  \\
\hline\hline
\end{tabular}
\end{table}

\clearpage

\squeezetable
\begin{table}
\caption{Diagnostics for importance of nondynamical correlation\label{tab:diagnostics}}
\begin{tabular}{l|ccccccccc}
\hline\hline
 & \%TAE[SCF] & ${\cal T}_1$ & $D_1$ & Largest T$_2$ & \%TAE[(T)] & \%TAE & \%TAE[$T_4+T_5$] & \multicolumn{2}{c}{NO occupations}  \\
 &  & \multicolumn{2}{c}{diagnostic} & amplitudes & & {}[post-CCSD(T)] &   & HDOMO & LUMO  \\
 &  &  \multicolumn{3}{c}{~~~~~~~~--- CCSD(T)/cc-pVTZ ---} &  &  &  & (a) &  \\
\hline
H$_2$O & 68.8 & 0.007          &  0.011  & 0.048 & 1.52 & -0.005 & 0.082   & 1.962 & 0.026                            \\
B$_2$  & 30.3 & 0.039          &  0.071  & 0.286 & 14.98 & 2.445 & 2.178    & 1.830 & 0.111                            \\
C$_2$H$_2$ & 74.0 & 0.013      &  0.028  & 0.084 (x2) & 2.07 & 0.012 & 0.190   & 1.924 (x2) & 0.057 (x2)          \\
CH$_3$ & 79.1 & 0.005          &  0.009  & 0.036, 0.034($\times 2$) & 0.62 & 0.008 & 0.018 & 1.959 & 0.022           \\
CH$_4$ & 78.9 & 0.007          &  0.011  & 0.035 ($\times 2$) & 0.69 & -0.001 & 0.019 & 1.958 (x3) & 0.022 (x3)      \\
CH & 68.1 & 0.008              &  0.017  & 0.09112 ($\times 2$) & 1.05 & 0.160 & 0.040  & 1.940 & 0.020                \\
CO$_2$ & 66.3 & 0.018          &  0.047  & 0.063 (x2) & 3.57 & 0.015 & 0.281       & 1.948 (x2) & 0.054 (x2)           \\
CO & 70.0 & 0.019              &  0.039  & 0.067 (x2) & 3.10 & 0.038 & 0.252           & 1.945 (x3) & 0.053 (x3)      \\
F$_2$ & -81.5 & 0.011          &  0.029  & 0.169 & 19.84 & 1.461 & 2.354           & 1.904 & 0.097                    \\
HF & 70.9 & 0.007              &  0.012  & 0.040 & 1.52 & -0.013 & 0.080               & 1.967 & 0.025                \\
N$_2$ & 52.4 & 0.013           &  0.026  & 0.095 (x2) & 4.18 & 0.152 & 0.498        & 1.931 (x2) & 0.063 (x2)         \\
NH$_3$ & 68.2 & 0.006          &  0.010  & 0.036 & 1.31 & 0.011 & 0.057            & 1.959 & 0.025                    \\
N$_2$O & 35.2 & 0.020          &  0.048  & 0.086 (x2) & 6.99 & 0.245 & 0.804       & 1.928 (x2) & 0.074 (x2)          \\
NO & 36.0 & 0.021              &  0.051  & 0.113 & 6.24 & 0.240 & 0.609                & 1.944 & 0.060                \\
O$_2$ & 22.3 & 0.007           &  0.013  & 0.094 & 7.73 & 0.357 & 0.971             & 1.950 & 0.042                   \\
O$_3$ & -30.8 & 0.027          &  0.077  & 0.192 & 17.74 & 1.954 & 2.866           & 1.876 & 0.014                    \\
C$_2$ & 12.5 & 0.038           &  0.086  & 0.293 & 13.33 & 0.311 & 1.813            & 1.629 & 0.362                   \\
BN & -10.8 & 0.073             &  0.199  & 0.224 & 18.75 & -0.316 & 2.226             & 1.833 & 0.157                 \\
MgO & -62.0 & 0.051            &  0.106  & 0.213 & 21.96 & 1.118 & 2.636             & 1.843 & 0.147                  \\
BeO & 51.1 & 0.043             &  0.104  & 0.041 (x2) & 8.11 & 0.168 & 0.368          & 1.943 (x2) & 0.044 (x2)       \\
CN & 46.9 & 0.053              &  0.152  & 0.092 (x2),  & 5.85 & 0.519 & 0.716         & 1.926 & 0.063                \\
 &  &                          &         & 0.077 (x2) &  &  &     &  &         \\
NO$_2$ & 26.2 & 0.025          &  0.065  & 0.093 & 8.56 & 0.447 & 0.923  & 1.943 & 0.077                              \\
Cl$_2$ & 46.4 & 0.008          &  0.021  & 0.091 & 7.72 & 0.132 & 0.749  & 1.930 & 0.065                              \\
ClF & 25.1 & 0.011             &  0.031  & 0.091 & 8.40 & 0.298 & 0.769       & 1.937 & 0.066                         \\
CS & 60.8 & 0.025              &  0.049  & 0.092 ($\times 2$) & 5.67 & 0.245 & 0.604 & 1.920 (x2) & 0.069 (x2)        \\
H$_2$S & 73.0 & 0.009          &  0.016  & 0.046, 0.045,  & 1.22 & 0.027 & 0.080 & 1.943 & 0.039                      \\
 &  &                          &         & 0.044, 0.043 &  &  &   &  &        \\
HCl & 75.9 & 0.006             &  0.011  & 0.043, 0.040 ($\times 4$) & 1.38 & -0.009 & 0.093 & 1.951 & 0.035          \\
HOCl & 52.6 & 0.010            &  0.024  & 0.061 & 4.08 & 0.077 & 0.361 & 1.932 & 0.064                               \\
PH$_3$ & 71.6 & 0.013          &  0.021  & 0.044 ($\times 2$),  & 0.83 & 0.028 & 0.039 & 1.933 & 0.037 (x2)           \\
 &  &                          &         & 0.038 ($\times 1+2$), &  &  &  &  &      \\
 &  &                          &         & 0.037 &  &  &  &  &      \\
SO & 42.4 & 0.023              &  0.052  & 0.074 & 6.73 & 0.082 & 0.708 & 1.942 & 0.043                               \\
SO$_2$ & 47.1 & 0.021          &  0.056  & 0.091 & 6.11 & 0.163 & 0.655  & 1.929 & 0.084                              \\
OCS & 65.3 & 0.019             &  0.049  & 0.067 ($\times 2$) & 4.33 & 0.065 & 0.392  & 1.932 (x2) & 0.062 (x2)       \\
ClCN & 59.7 & 0.014            &  0.028  & 0.074 ($\times 2$) & 4.39 & 0.069 & 0.476 & 1.927 (x2) & 0.063 (x2)        \\
C$_2$H$_4$ & 77.2 & 0.011      &  0.032  & 0.119 & 1.32 & -0.001 & 0.081  & 1.916 & 0.064                         \\
H$_2$CO & 70.8 & 0.016         &  0.045  & 0.119 & 2.12 & 0.013 & 0.154   & 1.927 & 0.063                         \\
HNO & 41.6 & 0.015             &  0.043  & 0.110 & 4.92 & 0.241 & 0.515       & 1.912 & 0.086                     \\
\hline\hline
\end{tabular}
\begin{flushleft}
Percentages of the total atomization energy relate to nonrelativistic, clamped-nuclei values with inner shell electrons
constrained to be doubly occupied.\\

$D_1$ diagnostics were obtained using MOLPRO 2006.1\cite{molpro}.\\

(a) Highest doubly occupied molecular orbital.\\

\end{flushleft}
\end{table}
\clearpage


\begin{thebibliography}{99}
\bibitem{H2footnote} As the H$_2$ molecule trivially has no post-CCSD(T) --- or for that matter post-CCSD --- correlation
contributions we have not included it in our statistics. For the sake of completeness, and in order to assist people 
wishing to obtain isodesmic or isogyric reaction energies involving H$_2$, we are providing the data here:
$D_0$[W2.2]=$D_0$[W3.2]=103.28, $D_0$[W4lite]=$D_0$[W4]=103.29 kcal/mol, compared to expt. value of 103.27 kcal/mol.
W4 component breakdown: TAE[SCF]=83.85, $\Delta$TAE[CCSD]=25.63, $\Delta$TAE[REL]=0.00, $\Delta$TAE[DBOC]=0.05, $D_e$=109.53 kcal/mol. 
Note that here the highest angular momentum actually present in the molecule was used for $L$ in the extrapolations: the other choice (cardinal number $N$ of the AV$N$Z basis sets, which is nonequivalent for H$_2$) would lead to different values of W2.2=W3.2=103.34 and W4lite=W4=103.32 kcal/mol, 
with the W4 SCF and valence correlation components being 83.85 and 25.65 kcal/mol, respectively.
The zero-point energy, 6.23 kcal/mol, was taken from 
spectroscopic constants in Huber and Herzberg\cite{Hub79}, with $Y_{00}$ being included.  (Failure to do so causes an underestimate of 0.025 kcal/mol.)
For H$_2$, there is\cite{ValeevSherrill} a nontrivial correlation correction to the DBOC of -0.04 kcal/mol: applying it would bring the W4 value to 103.25 kcal/mol with the former, and 103.28 kcal/mol with the latter, choice of $L$ in the extrapolations.
\bibitem{SOandSO2} As SO$_2$ has a central 2nd-row atom in a high oxidation state, it becomes quite vulnerable to `inner polarization' effects --- specifically, as discussed in Ref.\cite{cl2o7tae} for the extreme cases of HClO$_4$ and Cl$_2$O$_7$, the $(3d)$ orbitals drop low enough that they can accept some back-donation from lone pairs of highly electronegative elements. While the aug-cc-pV($n$+d)Z
basis set sequence is intended to cope with this, one may wonder if full saturation of the SCF component has in fact been achieved for SO$_2$. 
Using the AV$n$Z+2d1f sequence recommended in Ref.\cite{so2} --- specifically the AV5Z+2d1f and AV6Z+2d1f basis sets --- we obtain an SCF limit 0.05 kcal/mol higher for SO$_2$, and 0.02 kcal/mol higher for SO. In fact, in both cases this brings the W4 values closer to ATcT, although this may well be fortuitous.
\end{thebibliography}
\end{document}